\documentclass[11pt]{article}
\usepackage[letterpaper, left=.5in, top=0in, right=.5in, bottom=0in,nohead,includefoot, textwidth=4.5in]{geometry}
\usepackage{color,charter,graphicx,natbib,lscape,here,geometry,setspace,multirow,amsmath,amssymb,amsfonts,enumitem}
\usepackage[dvipsnames]{xcolor}
\usepackage{graphicx}
\usepackage{rotating}
\usepackage{multirow}
\usepackage{booktabs}
\usepackage{color}
\usepackage{setspace}
\usepackage{algorithm2e}
\usepackage{enumitem}

\usepackage[colorlinks=true,urlcolor=blue,linkcolor=blue,citecolor=blue]{hyperref}
\usepackage{bigstrut}
\usepackage{threeparttable}
\usepackage{dcolumn}
\usepackage{amsmath}
\usepackage{longtable}
\usepackage{amssymb}
\usepackage{bm}
\usepackage{comment}

\usepackage{mathtools}

\usepackage[hang,flushmargin]{footmisc} 
\usepackage[yyyymmdd]{datetime}
\usepackage[british]{babel}

\pdfoutput=1
 \usepackage{float}
 \restylefloat{table}

\usepackage[title,titletoc]{appendix}
\makeatletter

\renewenvironment{appendices}{%
    \begin{oldappendices}%
    \renewcommand{\thefigure}{\ifnum \c@section>\z@ \thesection.\fi\@arabic\c@figure}%
    \@addtoreset{figure}{section}%
    \renewcommand{\thetable}{\ifnum \c@section>\z@ \thesection.\fi\@arabic\c@table}%
    \@addtoreset{table}{section}%
}{%
    \end{oldappendices}%
}\makeatother

\usepackage{natbib}
\let\natbibcitet\citet
\renewcommand\citet{\bibpunct{(}{)}{,}{a}{,}{,}\natbibcitet}
\let\natbibcitep\citep
\renewcommand\citep{\bibpunct{(}{)}{;}{a}{,}{;}\natbibcitep}
\newcommand{\bi}{\begin{itemize}}
\newcommand{\ei}{\end{itemize}}
\newcommand{\be}{\begin{equation}}
\newcommand{\ee}{\end{equation}}
\defcitealias{ieo14}{IEO, 2014}

\long\def\symbolfootnote[#1]#2{\begingroup%
\def\thefootnote{\fnsymbol{footnote}}\footnote[#1]{#2}\endgroup}

\usepackage{sectsty}
\allsectionsfont{\normalsize}

\makeatletter
\def\ubar#1{\underline{\sbox\tw@{$#1$}\dp\tw@\z@\box\tw@}}
\makeatother

\usepackage{bm}
\usepackage{caption}
\usepackage{subcaption}

\makeatletter\let\p@subfigure\thefigure\makeatother

\usepackage{cleveref}
\crefname{chapter}{Chapter}{Chapters}
\crefname{section}{Section}{Sections}
\crefname{subsection}{Section}{Sections}
\crefname{subsubsection}{Section}{Sections}
\crefname{figure}{Figure}{Figures}
\crefname{table}{Table}{Tables}
\crefname{equation}{Equation}{Equations}
\crefname{appendix}{Appendix}{Appendices}
\crefname{appendices}{Appendix}{Appendices}

\crefname{appsec}{Appendix}{Appendices}

\usepackage{catoptions}
\makeatletter

\def\Autoref#1{%
  \begingroup
  \edef\reserved@a{\cpttrimspaces{#1}}%
  \ifcsndefTF{r@#1}{%
    \xaftercsname{\expandafter\testreftype\@fourthoffive}
      {r@\reserved@a}.\\{#1}%
  }{%
    \ref{#1}%
  }%
  \endgroup
}
\def\testreftype#1.#2\\#3{%
  \ifcsndefTF{#1autorefname}{%
    \def\reserved@a##1##2\@nil{%
      \uppercase{\def\ref@name{##1}}%
      \csn@edef{#1autorefname}{\ref@name##2}%
      \autoref{#3}%
    }%
    \reserved@a#1\@nil
  }{%
    \autoref{#3}%
  }%
}
\makeatother

\newcolumntype{d}[1]{D{.}{.}{#1}}

\title{The dynamic impact of monetary policy on regional housing prices in the US: Evidence based on factor-augmented vector autoregressions}
\date{}
\usepackage{authblk}

\author{Manfred M. Fischer, Florian Huber, Michael Pfarrhofer and Petra Staufer-Steinnocher\\
Vienna University of Economics and Business}


\def\equationautorefname~#1\null{%
  Eq.~(#1)\null
}
\def\equationautorefname~#1\null{
Eq.~(#1)\null
}

\usepackage[utf8, latin1]{inputenc}                 

\begin{document}
\clearpage\maketitle
\thispagestyle{empty}

\onehalfspacing

\begin{abstract}
\noindent In this study interest centers on regional differences in the response of housing prices to monetary policy shocks in the US. We address this issue by analyzing monthly home price data for metropolitan regions using a factor-augmented vector autoregression (FAVAR) model. Bayesian model estimation is based on Gibbs sampling with Normal-Gamma shrinkage priors for the autoregressive coefficients and factor loadings, while monetary policy shocks are identified using high-frequency surprises around policy announcements as external instruments. The empirical results indicate that monetary policy actions typically have sizeable and significant positive effects on regional housing prices, revealing differences in magnitude and duration. The largest effects are observed in regions located in states on both the East and West Coasts, notably California, Arizona and Florida.
\end{abstract}

\bigskip
\begin{tabular}{p{0.2\hsize}p{0.65\hsize}} 
\textbf{Keywords:}  &Regional housing prices, metropolitan regions, Bayesian estimation, high-frequency identification\\
\end{tabular}

\smallskip
\begin{tabular}{p{0.2\hsize}p{0.4\hsize}}
\textbf{JEL Codes:} &C11, C32, E52, R31 \\
\end{tabular}
\vspace{0.6cm}
 
\begin{center}
\date{\today}
\end{center}

\bigskip
\newpage
\section{Introduction}
This paper examines the impact of monetary policy on housing prices in the US.\footnote{Housing is defined here to include family residences, condominiums and co-operative homes.} The literature on this relationship is fairly limited. Previous studies generally rely on two competing approaches. The first uses a structural model to analyze the relationship between monetary policy and housing prices (see, for example, \citealp{MANC:MANC332,ungerer2015monetary}). Such models impose a priori restrictions on the coefficients. The major strength of this model-based approach is to provide a theoretically grounded answer to the question of interest. Its potential shortcoming, however, is that the answer is only as good as the model is adequately representing the relationships in the real world.

The second approach -- labeled evidence-based -- focuses more on the empirical evidence and relies less directly on economic theory. Researchers have commonly used vector autoregressive (VAR) models to measure the impact of monetary policy \citep[see][]{Baffoe-Bonnie1998,fratatoni2003monetary,delnegro2007luftballons,jarocinski2008house,vargassilva2008monetary,beltratti2010international,ECTJ:ECTJ319}. Such models allow the data rather than the researcher to specify the dynamic structure of the model, and provide a plausible assessment of the response of macroeconomic variables to monetary policy shocks without the need of a complete structural model of the economy.

In the tradition of the latter approach, this paper differs from previous literature both in terms of focus and methodology. With \cite{fratatoni2003monetary}, we share the focus on regional differences in the response of housing prices, using metropolitan-level rather than state-level data.\footnote{Their empirical analysis uses a small set of 27 US regions to analyze the effects of monetary policy, based on quarterly data from 1986 to 1996. Aside from this study, metropolitan-level housing data have not been explored very much.} In terms of methodology, similar to \cite{vargassilva2008monetary} and in contrast to \cite{fratatoni2003monetary}, we use a factor-augmented vector autoregressive (FAVAR) model to explore regional housing price responses to a national monetary policy shock.\footnote{For the definition of our notion of region and the list of regions used, see \Cref{app:regions}.} The effects are measured by considering idiosyncratic impulse responses of regions to the shock that is normalized to yield a 25 basis-points decline in the one-year government bond rate.

Differently from \cite{vargassilva2008monetary} and \cite{ECTJ:ECTJ319}, we employ a full Bayesian approach that is based on shrinkage priors for several parts of the parameter space. In particular, we make use of Markov Chain Monte Carlo (MCMC) methods to estimate the model parameters and the latent factors simultaneously. A full Bayesian approach has the advantage of directly controlling for uncertainty surrounding the latent factors and the model parameters. We follow \cite{gertler2015monetary} to identify monetary policy shocks by using high-frequency surprises around policy announcements as external instruments.

The paper provides a rich picture on how an expansionary monetary policy shock affects housing prices in 417 US metropolitan regions over a time horizon of 72 months after impact. The findings reveal regional housing price effects to vary substantially over space, with size and modest sign differences among the regions. Some few regions in Utah, New Mexico, Kansas, Oklahoma, Mississippi and West Virginia show no significant impact or even slightly negative cumulative responses. In most of the regions, however, the cumulative responses of housing prices are positive, in line with theory. This regional heterogeneity may have different reasons, with heterogeneous regional housing markets playing a major role. The largest positive effects are observed in states on both the East and West Coasts, notably in Miami-Fort Lauderdale in Florida and Riverside-Sun Bernardino-Ontario in California, but also in Las Vegas in Nevada. In general, housing impulse responses tend to be similar within states and adjacent regions in neighboring states, evidenced by a high degree of spatial autocorrelation.

The remainder of the paper is structured as follows. The next section presents the FAVAR model along with the Bayesian approach for estimation. \Autoref{sec:dataimplementation} describes the data and the sample of regions, and outlines the model specification. The empirical results are discussed in \autoref{sec:results}, while the final section concludes.

\section{Econometric framework}\label{sec:framework}
\subsection{A factor-augmented vector autoregressive model}
The econometric approach employed in this study is a FAVAR model, as introduced in \cite{bernanke2005measuring}. In our implementation, we let  $\bm{H}_t=(H_{1t},...,H_{Rt})'$ denote an $R \times 1$ vector of housing prices at time $t$ ($t = 1, \hdots, T$) across $R=417$ US regions. The model postulates that regional housing prices depend on a number of latent factors, monetary and macroeconomic national aggregates and region-specific shocks. Specifically, the measurement equation can be written as
\begin{equation}
\begin{bmatrix} \bm{H}_t \\ \bm{M}_t  \end{bmatrix} = 
\begin{bmatrix*}[l]
\bm{\Lambda}^F & \bm{\Lambda}^M \\ 
\bm{0}_{K \times S} & \bm{I}_K
\end{bmatrix*} 
\begin{bmatrix}
\bm{F}_t \\
\bm{M}_t
\end{bmatrix} + \begin{bmatrix*}[l]
\bm{\epsilon} _t \\
\bm{0}_{K \times 1}
\end{bmatrix*},
\label{eq:main-struct}
\end{equation}
where $\bm{F}_t=(F_{1t},...,F_{St})'$ is an $S \times 1$ vector of latent (unobservable) factors which capture co-movement at the regional level ($F_{rt}$, $r = 1, \hdots, R$, $t = 1, \hdots, T$). $\bm{M}_t=(M_{1t},...,M_{Kt})'$ is a $K \times 1$ vector of economic and monetary national aggregates that are treated as observable factors, and $\bm{\epsilon}_t$ ($t = 1, \hdots, T$) an $R \times 1 $ vector of normally distributed zero mean disturbances with an $R \times R$ variance-covariance matrix $\bm{\Sigma}_{\epsilon}=\text{diag}(\sigma_1^2, \dots, \sigma_R^2)$. These disturbances arise from measurement errors and special features that are specific to individual regional time series. $\bm{\Lambda}^F = (\lambda^{F}_{rs}:$ $r = 1, \hdots, R$; $s = 1, \hdots, S)$ is an $R \times S$ matrix of factor loadings with typical elements $\lambda^{F}_{rs}$, while $\bm{\Lambda}^M = (\lambda^{M}_{rk}:$ $r = 1, \hdots, R$; $k = 1, \hdots, K)$ a coefficient matrix of dimension $R \times K$ with typical elements $\lambda^{M}_{rk}$. The number of latent factors is much smaller than the number of regions, i.e. $S \ll R$. Note that the diagonal structure of $\bm{\Sigma}_\epsilon$ implies that any co-movement between the elements in $\bm{H}_t$ and $\bm{M}_t$ stems exclusively from the presence of the factors.

The evolution of the factors $\bm{y}_t=(\bm{F}'_t, \bm{M}'_t)'$ is given by the state equation, governed by a VAR process of order $Q$, 
\begin{equation}
\bm{y}_t =  \bm{A} \bm{x}_t + \bm{u}_t \label{eq: stateEQ},
\end{equation}
with $\bm{x}_t=(\bm{y}'_{t-1}, \dots, \bm{y}'_{t-Q})'$ and the associated $(S+K)\times Q(S+K)$-dimensional coefficient matrix $\bm{A}$. Moreover, $\bm{u}_t$ is an $(S+K)$-dimensional vector of normally distributed shocks, with zero mean and variance covariance matrix $\bm{\Sigma}_u$. 

The parameters $\bm{\Lambda}^F$, $\bm{\Lambda}^{M}$ and $\bm{A}$ as well as the latent dynamic factors $\bm{F}_{t}$ are unkown and have to be estimated. To identify the model, we follow \citet{bernanke2005measuring} and assume that the upper $(S\times S)$-dimensional submatrix of $\bm{\Lambda}^F$ equals an identity matrix $\bm{I}_S$ while the first $S$ rows of $\bm{\Lambda}^{M}$ are set equal to zero. This identification strategy implies that the first $S$ elements in $\bm{H}_t$ are effectively the factors plus noise.

\subsection{A Bayesian approach to estimation}
The model described above is highly parameterized, containing more parameters that can be reasonably estimated with the data at hand. In this study, we use a Bayesian estimation approach to incorporate knowledge about parameter values via prior distributions. Before proceeding with the prior setting employed it is convenient to stack the free elements of the factor loadings in an $L$-dimensional vector $\bm{\lambda}=\text{vec}[\bm{\Lambda}^F, \bm{\Lambda}^M]$  with $L=R (S+K)$, and the VAR coefficients in a $J$-dimensional vector $\bm{a}=\text{vec}(\bm{A})$ with $J=(S+K)^2 Q$.

\subsection*{Prior distributions for the state equation}
For the VAR coefficients $a_j$ ($j = 1,\hdots,J$) we impose the Normal-Gamma shrinkage prior proposed in \citet{griffin2010inference,griffin2016hierarchical}, and subsequently applied in a VAR framework in \citet{Huber2017},
\begin{equation}
a_j| \xi_a, \tau_{a j}^2 \sim \mathcal{N}\left(0, 2\ \xi_a^{-1} \tau^2_{a j}\right),
\end{equation}
that is controlled by Gamma priors on $\tau^2_{a j}~(j = 1, \hdots, J)$ and $\xi_a$,
\begin{align}
\tau^2_{a j} \sim \mathcal{G}(\vartheta_a, \vartheta_a), \label{eq:tausq}\\
\xi_a \sim \mathcal{G}(d_0, d_1), \label{eq:xi_a}
\end{align}
with hyperparameters $\vartheta_a$ and $d_0$, $d_1$ respectively. $\tau^2_{a j}$ operates as a local scaling and $\xi_a$ as a global shrinkage parameter.

This hierarchical prior shows two convenient features. \textit{First}, $\xi_a$ applies to all $J$ elements in $\bm{a}$. Higher values of $\xi_a$ yield stronger global shrinkage towards the origin whereas smaller values induce only little shrinkage. \textit{Second}, the local scaling parameters $\tau^2_{a j}$ place sufficient prior mass of $a_j$ away from zero in the presence of strong overall shrinkage involved by large values for $\xi_a$.

The hyperparameter $\vartheta_a$ in \autoref{eq:tausq} controls the excess kurtosis of the marginal prior,
\begin{equation}
p(a_j | \xi_a) = \int p(a_j| \xi_a, \tau_{a j}^2 ) d\tau^2_{a j},
\end{equation}
obtained after integrating over the local scales. Lower values of $\vartheta_a$ generally place increasing mass on zero, but at the same time lead to heavy tails, allowing for large deviations of $a_j$ from zero, if necessary. The hyperparameters $d_0$ and $d_1$ in \autoref{eq:xi_a} are usually set to rather small values to induce heavy overall shrinkage \citep[see][for more details]{griffin2010inference}.

For the variance-covariance matrix $\bm{\Sigma}_u$ we use an inverted Wishart prior,
\begin{equation}
\bm{\Sigma}_u \sim \mathcal{IW}(v, \bm{\ubar{\Sigma}}),
\end{equation}
with $v$ denoting prior degrees of freedom, while $\bm{\ubar{\Sigma}}$ is a prior scaling matrix of dimension $(S+K)\times(S+K)$.

\subsection*{Prior distributions for the observation equation}
For the factor loadings $\lambda_{\ell}$ ($\ell = 1, \hdots, L$) we employ a Normal-Gamma prior similar to the one used for the VAR coefficients $a_j$ ($j = 1, \hdots, J$). The set-up follows \citet{kastner} with a single global shrinkage parameter $\xi_\lambda$ that applies to all free elements $\lambda_\ell$ in the factor loadings matrix. Specifically, we impose a hierarchical Gaussian prior on $\lambda_\ell$,
\begin{equation}
\lambda_\ell|\xi_\lambda, \tau^2_{\lambda \ell} \sim \mathcal{N}\left(0, 2\ \xi_\lambda^{-1} \tau^2_{\lambda \ell}\right)
\end{equation}
that depends on Gamma priors for $\tau^2_{\lambda \ell}~(\ell = 1, \hdots, L)$ and $\xi_\lambda$,
\begin{align}
\tau^2_{\lambda\ell} \sim \mathcal{G}(\vartheta_\lambda, \vartheta_\lambda),\\
\xi_\lambda \sim \mathcal{G}(c_0, c_1).
\end{align}
The hyperparameters $\vartheta_\lambda, c_0$ and $c_1$ control the tail behavior and overall degree of shrinkage of the prior. 

For the measurement error variances $\sigma^2_r~(r = 1,\hdots,R)$ we rely on a sequence of independent inverted Gamma priors,
\begin{equation}
\sigma_r^2 \sim \mathcal{G}^{-1}(e_0, e_1),
\end{equation}
where the hyperparameters $e_0$ and $e_1$ are typically set to small values to reduce prior influence on $\sigma^2_r$. 

Estimation of the model parameters and the latent factors is based on the MCMC algorithm described in \Cref{app:mcmc}. More specifically, we use Gibbs sampling to simulate a chain consisting of 20,000 draws, where we discard the first 10,000 draws as burn-in. It is worth noting that this algorithm shows fast mixing and satisfactory convergence properties.

\section{Data and model implementation}
\label{sec:dataimplementation}
\subsection{Regions and Data}
To explore regional differences in the impact of monetary policy on housing prices in the US, we need to define our notion of regions. Throughout the paper, we use $R = 417$ regions, a subsample of the 917 core-based statistical areas.\footnote{A core-based statistical area is a US geographic area -- defined by the Office of Management and Budget -- that consists of one or more counties anchored by an urban center of at least 10,000 people plus adjacent counties that are socioeconomically tied to the urban center. The term core-based statistical area refers collectively to both metropolitan and micropolitan statistical areas.} These 417 regions include 264 metropolitan and 153 micropolitan statistical areas, briefly termed metropolitan regions in this paper. They have been selected based on the availability of the data over time. For the list of regions used, see \Cref{app:regions}.

Our dataset consists of a panel of monthly time series ranging from 1997:04 to 2012:06. The $R \times 1$ vector of housing prices at time $t$ is constructed using the Zillow Home Value Index.\footnote{The Zillow Home Value Index has the benefit of a broad coverage of the large set of core-based statistical areas. The set of data we use in our study is available for download at \url{https://www.zillow.com/research/data/}.} In contrast to the FHFA (Federal Housing Finance Agency) Index and the Standard \& Poor's Case-Shiller Index, the Zillow Home Value Index does not use a repeat sales methodology, but statistical models along with information from sales assessments to generate valuations for all homes (single family residences, condominiums and co-operative homes) in any given region.\footnote{For more information on the proprietary valuation model used by Zillow to estimate the market value of a home, see \citet{bruce2014zillow}.} These valuations are aggregated to determine the Zillow Home Value Index, measured in US dollars. An estimate for any given property is meant to indicate the fair value of a home sold as a conventional non-foreclosure, arms-length sale \citep{winkler2013}.

We include $K = 7$ variables in the $K \times 1$ vector of observable national aggregates: three economic variables, namely housing investment (measured as the quantity of housing starts), the industrial production index and the consumer price index. The one-year government bond rate serves as policy indicator in line with \citet{gertler2015monetary}. In addition, three credit-spreads are included: the ten-year treasury minus the federal funds rate, the prime mortgage spread calculated over the ten-year government bonds and the \citet{gilchrist2012credit} excess bond premium.\footnote{The excess bond premium may roughly be seen as the component of the spread between an index of yields on corporate fixed income securities and a similar maturity government bond rate that is left after removing the component due to default risk \citep{gertler2015monetary}. \citet{gilchrist2012credit} show that this variable provides a convenient summary of additional information not included in the FAVAR that may be relevant to economic activity.} The economic variables capture housing, price and output movements. The mortgage spread is relevant to the cost of housing finance and the excess bond premium to the cost of long term credit in the business sector, while the term spread measures expectations on short-term interest rates \citep{gertler2015monetary}. All observable national aggregates are taken from the FRED database \citep{McCracken2016}, with the exception of the excess bond premium and the mortgage spread that we obtained from the dataset provided in \citet{gertler2015monetary}. All data series are seasonally adjusted, if applicable, and transformed to be approximately stationary. 

\subsection{Model implementation}
For implementation of the FAVAR, we have to specify the order $Q$ of the VAR process and the number of latent factors, $S$. As is standard in the literature, we pick $Q = 2$ lags of the endogenous variables. To decide on the number of factors we use the deviance information criterion \citep{spiegelhalter2002bayesian} where the full data likelihood is obtained by running the Kalman filter and integrating out the latent states. This procedure yields $S=1$, a choice that is also consistent with traditional criteria (Bayesian information criterion or Kaiser criterion) to select the number of factors.

Next and finally, a brief word on hyperparameter selection for the prior set-up. We specify $\vartheta_a = \vartheta_\lambda=0.1$, a choice that yields strong shrinkage but, at the same time, leads to heavy tails in the underlying marginal prior. Recent literature \citep{Huber2017} integrates out $\vartheta_a, \vartheta_\lambda$ and finds that, for US data, the posterior is centered on values between $0.10$ and $0.15$. The hyperparameters on the global shrinkage parameters are set equal to $c_0=c_1= d_0 =d_1 = 0.01$, a choice that is consistent with heavy shrinkage towards the origin representing a standard in the literature \citep{griffin2010inference}. The prior on $\bm{\Sigma}_u$ is specified to be weakly informative, i.e. $\nu= S+K+1$ and $\bm{\ubar{\Sigma}}= 10^{-2} \bm{I}_{S+K}$. Likewise, for the inverted Gamma prior on $\sigma_r^2~(r = 1, \hdots, R)$ we set $e_0=e_1=0.01$ to render the prior only weakly influential.

\section{Impulse response analysis}\label{sec:results}
\subsection{Structural identification of the model}
The high-frequency variant of the external instruments identification approach \citep{kuttner2001monetary, gurkaynak2005sensitivity} employed in this paper is based on the surprises in the three-months-ahead futures rate that reflect expectations on interest rate movements further into the future, measured within a 30 minutes time window surrounding Federal Reserve announcements \citep{gertler2015monetary}. Note that in contrast to the Cholesky identification strategy, there is no need to impose zero restrictions.

To implement the approach we follow \citet{paul20017} and use high-frequency surprises as a proxy for the structural monetary policy shock. This is achived by integrating the surprises into \autoref{eq: stateEQ} as an exogenous variable ${z}_t$,
\begin{equation}
\bm{y}_t = \bm{A} \bm{x}_t + \bm{\zeta} {z}_t+ \bm{u}_t.
\end{equation}
Hereby $\bm{\zeta}$ is a $Q(S+K)$-dimensional vector of regression coefficients that collects the impulses of the shocks. \citet{paul20017} shows that under mild conditions, the contemporaneous relative impulse responses can be estimated consistently.\footnote{Relative impulse responses are obtained by normalizing the absolute impulse responses, i.e. the change in $\bm{y}_{t+h}$ to a change in $z_t$, by the contemporaneous response of some element in $\bm{y}_{t}$. } Note that the contemporaneous  response of $\bm{y}_t$ to changes in $z_t$ is given by $\bm{\zeta}$.  Higher order responses are defined recursively by exploiting the state space representation of the VAR model in \autoref{eq: stateEQ}.

\subsection{Impulse responses of macroeconomic quantities}
We first consider the dynamic responses of the endogenous variables included in the $K \times 1$ vector $\bm{M}_t~(t = 1, \hdots, T)$ to illustrate that the results of the model are consistent with established findings in the literature. An expansionary monetary policy shock is modeled by taking the one-year government bond rate as the relevant policy indicator, rather than the federal funds rate that is commonly used in the literature based on arguments presented in \citet{gertler2015monetary}.\footnote{\citet{gertler2015monetary} have shown that the one-year bond rate has a much stronger impact on market interests than the funds rate does.} Normalization is achieved by assuming that a monetary policy shock yields a 25 basis-points decrease in the policy indicator.

\begin{figure}[!t]
\centering
\begin{subfigure}{.329\textwidth}
\subcaption{Industrial production index}\label{fig:ip}
\includegraphics[width=\textwidth]{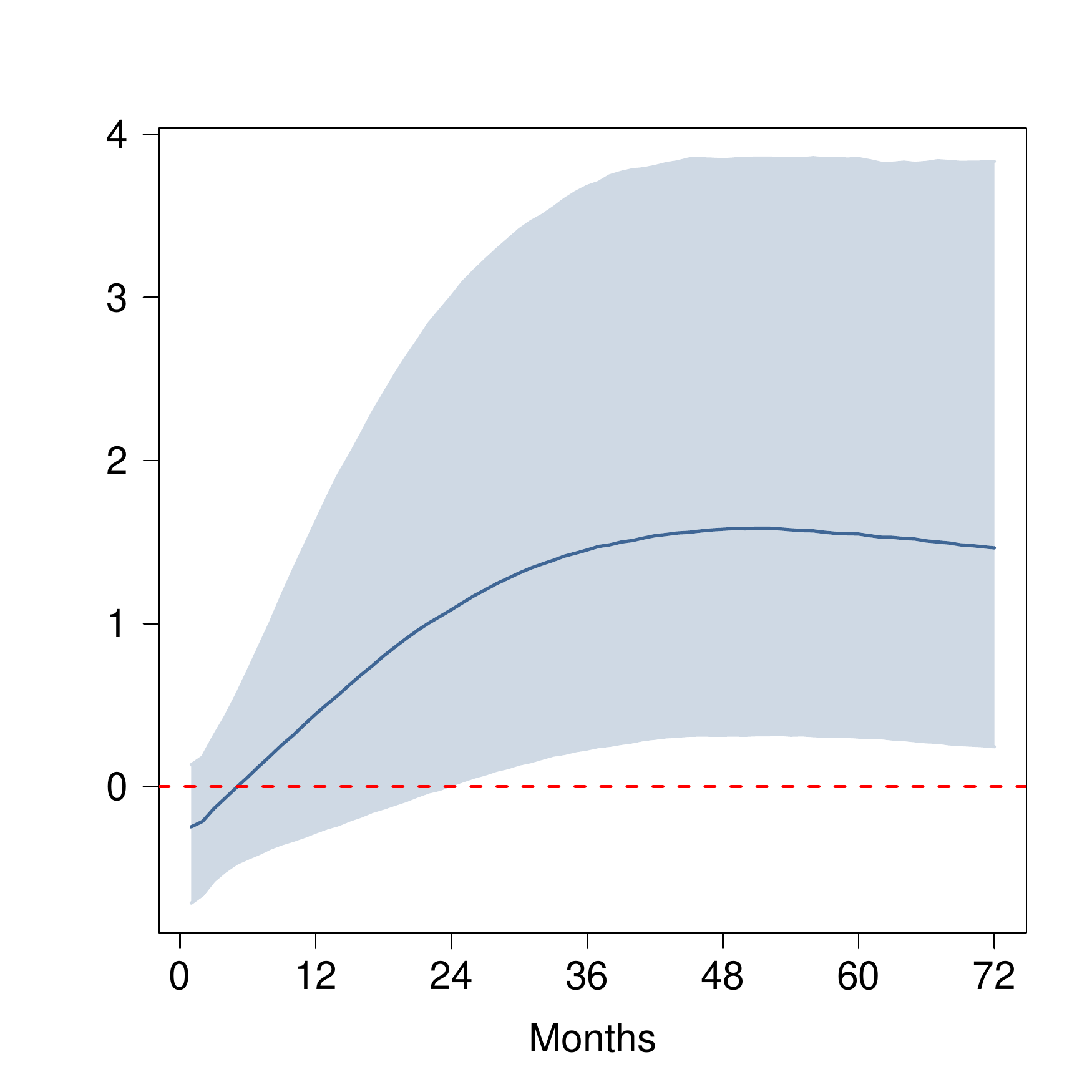}
\end{subfigure}
\begin{subfigure}{.329\textwidth}
\subcaption{Housing investment}\label{fig:houst}
\includegraphics[width=\textwidth]{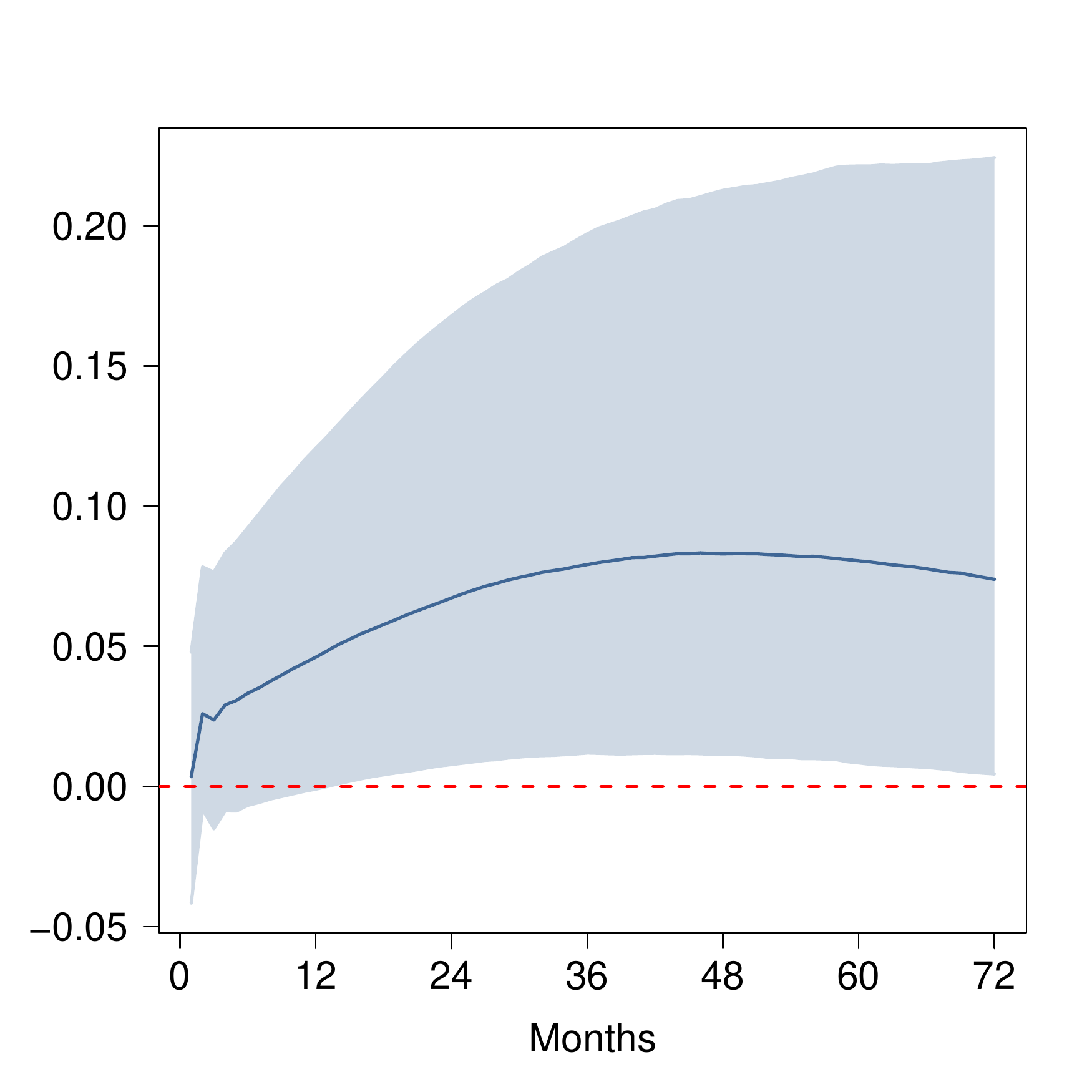}
\end{subfigure}
\begin{subfigure}{.329\textwidth}
\subcaption{Consumer price index}\label{fig:cpi}
\includegraphics[width=\textwidth]{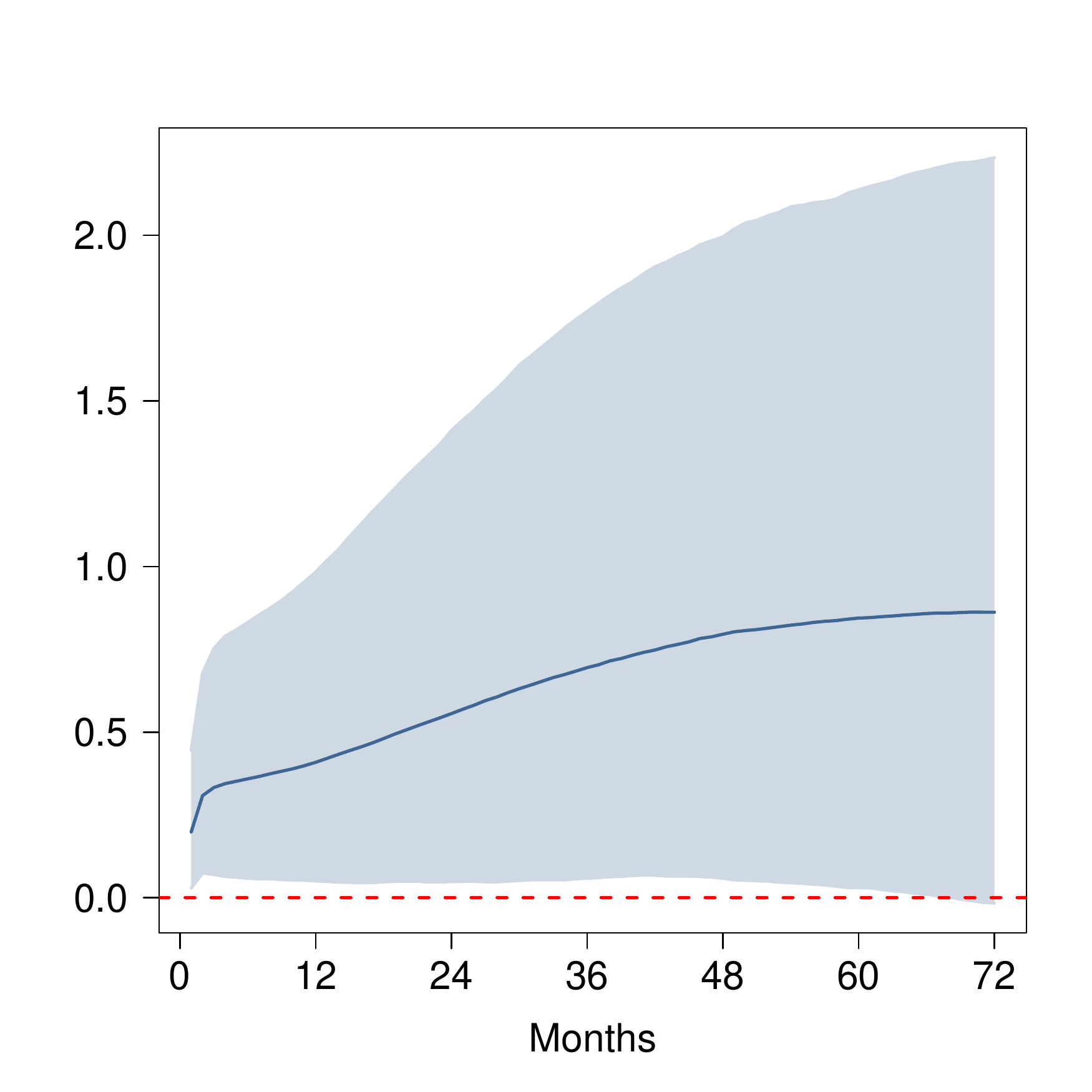}
\end{subfigure}\\

\begin{subfigure}{.329\textwidth}
\subcaption{One-year government bond rate}\label{fig:t10yty}
\includegraphics[width=\textwidth]{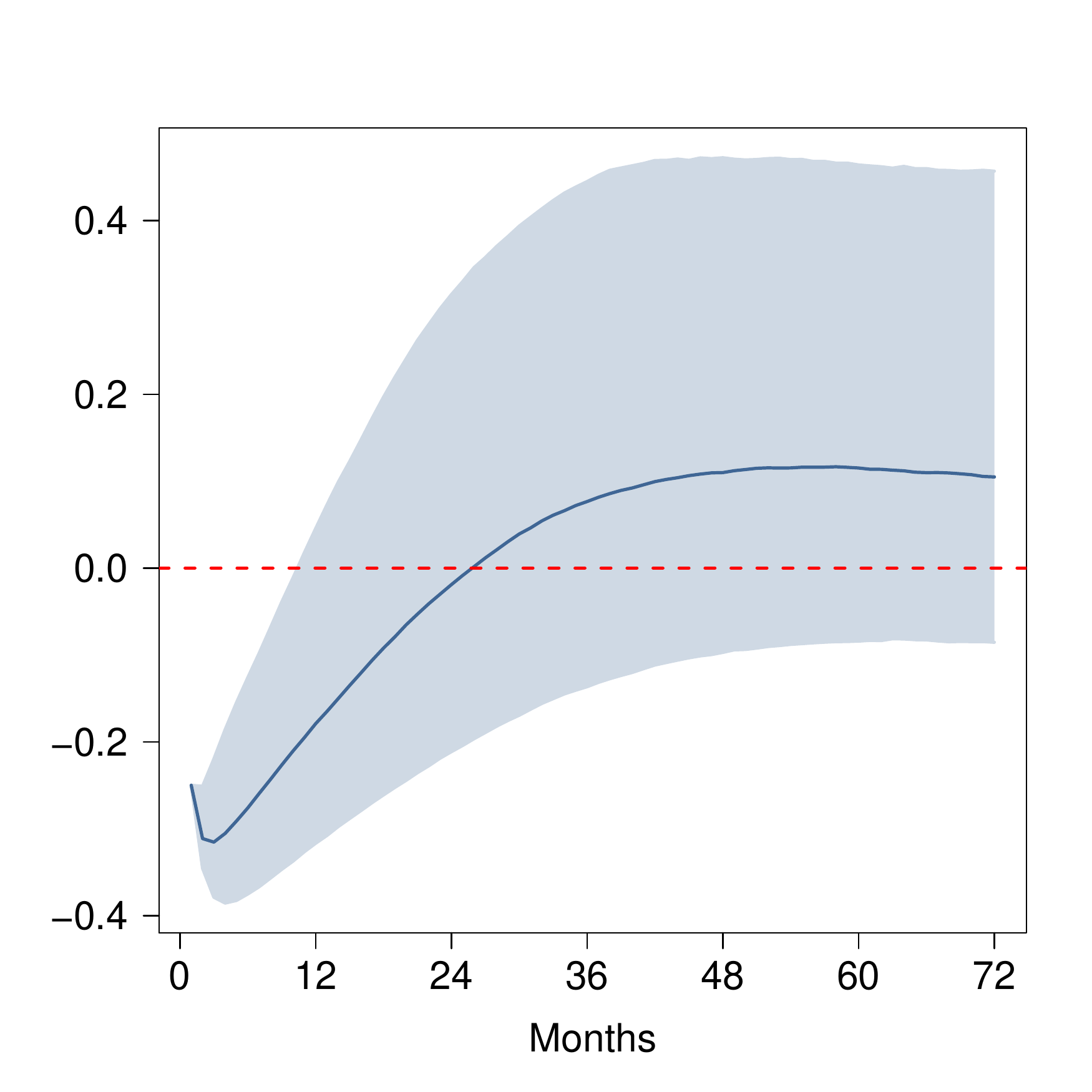}
\end{subfigure}
\hfill
\begin{subfigure}{.329\textwidth}
\subcaption{Term spread}\label{fig:termspread}
\includegraphics[width=\textwidth]{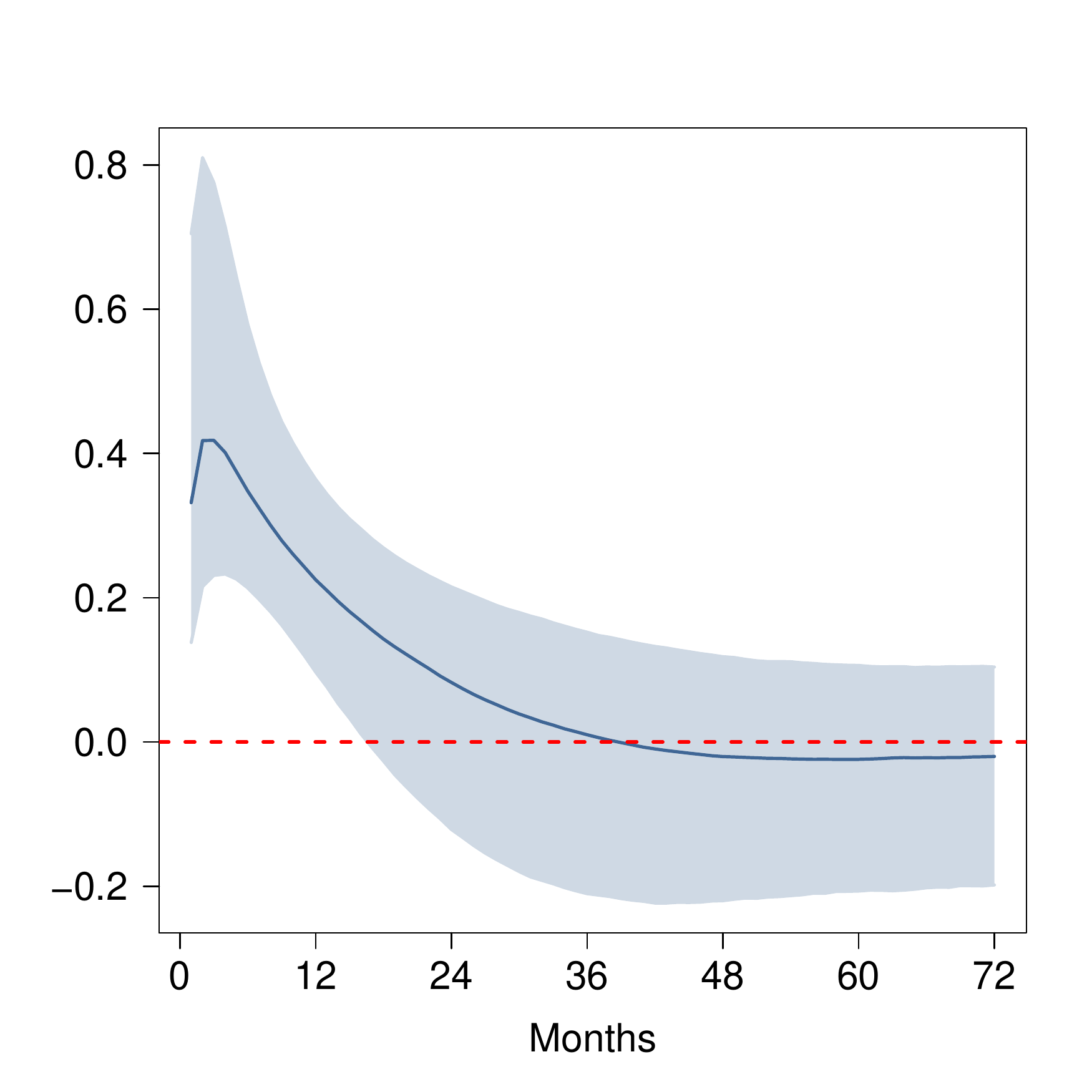}
\end{subfigure}
\begin{subfigure}{.329\textwidth}
\subcaption{Prime mortgage spread}\label{fig:mortgage_spread}
\includegraphics[width=\textwidth]{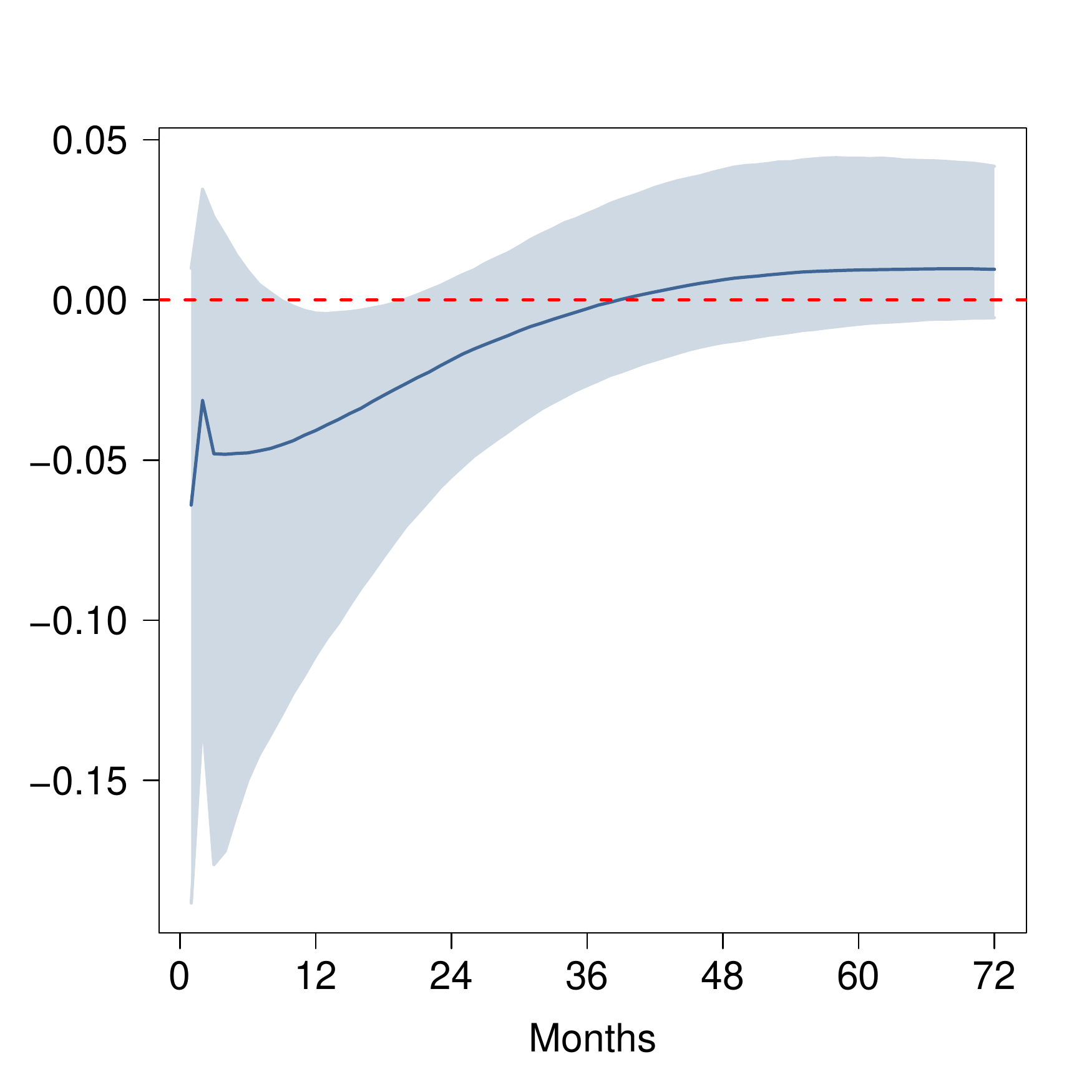}
\end{subfigure}\\

\begin{subfigure}{.329\textwidth}
\subcaption{Excess bond premium}\label{fig:ebp}
\includegraphics[width=\textwidth]{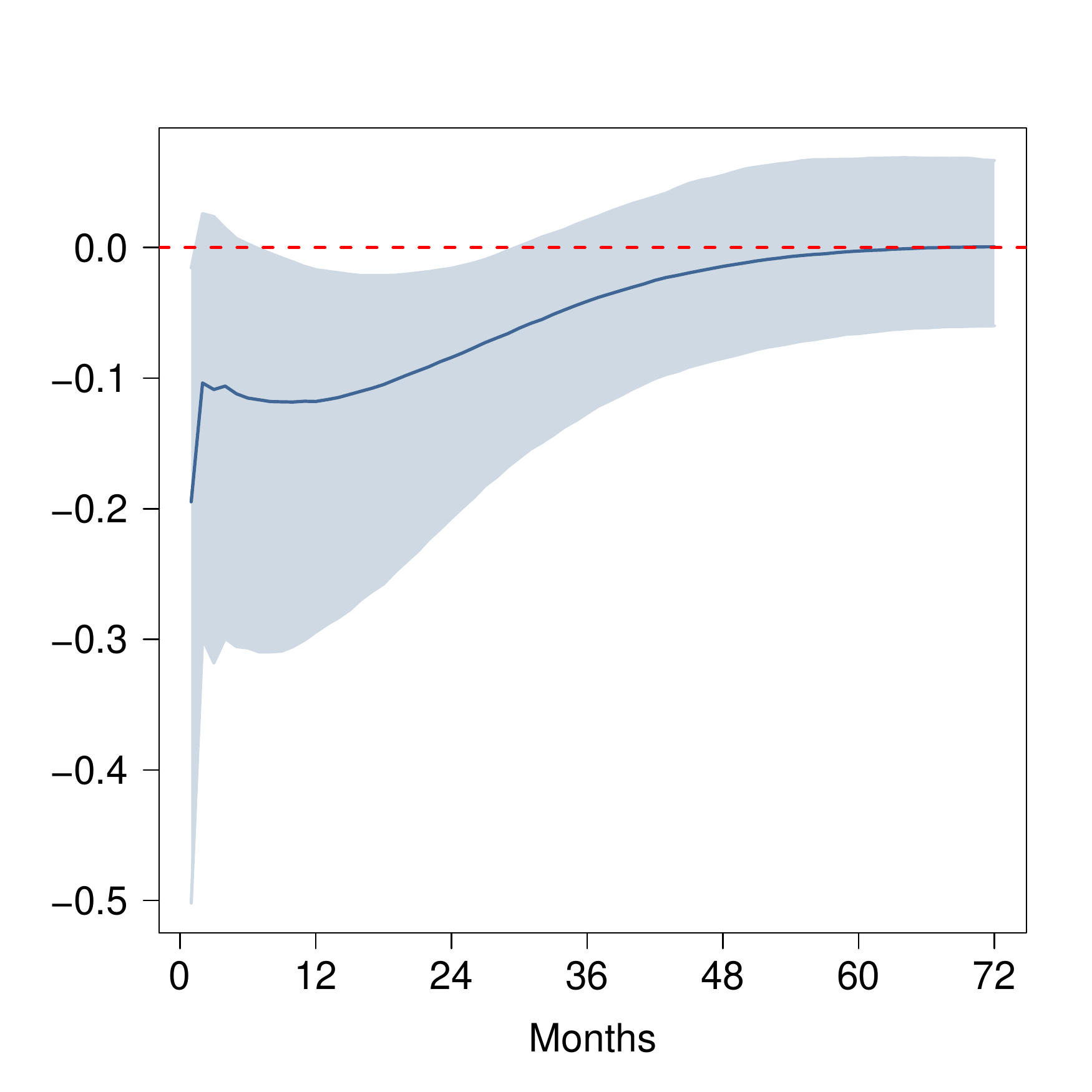}
\end{subfigure}\\
\caption{Impulse responses of macroeconomic fundamentals to a monetary policy shock.}
\caption*{\footnotesize{\textit{Notes}: The solid blue line denotes the median response, the dashed red line the zero line, and the shaded bands (in light blue) the 68 percent posterior coverage interval. Results are based on 10,000 posterior draws. Sample period: 1997:04 -- 2012:06. Vertical axis: percentage changes for indices and housing investment; otherwise percentage points. Front axis: months after impact.}}
\label{fig:IRF_macro}
\end{figure}

\cref{fig:IRF_macro} depicts the impulse response functions of the endogenous variables. All plots include the median response (in blue) for 72 months after impact along with 68 percent posterior coverage intervals reflecting posterior uncertainty. An unanticipated decrease in the government bond rate by 25 basis-points causes a significant increase in real activity, with industrial production, housing investment and consumer prices all increasing over the next months after the impact. From a quantitative standpoint, the effects of the monetary shock on industrial production and consumer price index are considerably larger than the impact on housing investment, although uncertainty surrounding the size of impacts is large, and posterior coverage intervals include zero during the first months after impact. Housing investment shows a reaction similar in shape to real activity measured in terms of the industrial production index, suggesting a positive relationship between expansionary monetary policy and housing investment at the national level.

Turning to the responses of financial market indicators, it should be noted that the one-year government bond rate falls by 25 basis-points on impact by construction, then increases significantly before it turns non-significant after about nine months. The term spread reacts adversely on impact, and we find significant deviations from zero that die out after about 16 months. This result points towards an imperfect pass-through of monetary policy on long-term rates, implying that long-term yields display a weaker decline as compared to short-term rates. The prime mortgage spread does not show a significant effect on impact, while responses between ten to 20 months ahead indicate a slightly negative overall reaction to expansionary monetary policy. Consistent with \citet{gilchrist2012credit}, one implication of this finding is that movements in key short-term interest rates tend to impact credit markets, with mortage spreads showing a tendency to decline. The responses of the excess bond premium almost perfectly mirror the reaction of the mortgage spread. The effects, however, are much larger from a quantitative point of view.

To sum up, the results obtained by the impulse response analysis provide empirical support that monetary policy shocks, identified by using high-frequency surprises around policy announcements as external instruments, generate impulse responses of the endogenous variables that are consistent with the findings by \citet{gertler2015monetary}.

\subsection{The dynamic factor and its loadings}
Before moving to the impulse responses of regional housing markets to a monetary policy shock, we briefly consider the estimated latent factor as well as its loadings, with two aims in mind: first, to provide a rough intuition on how the latent factor captures co-movement in regional house price variations, and second, to give indication of the relative importance of individual regions shaping the evolution of the common factor.

\begin{figure}[H]
\centering
\includegraphics[scale=.55]{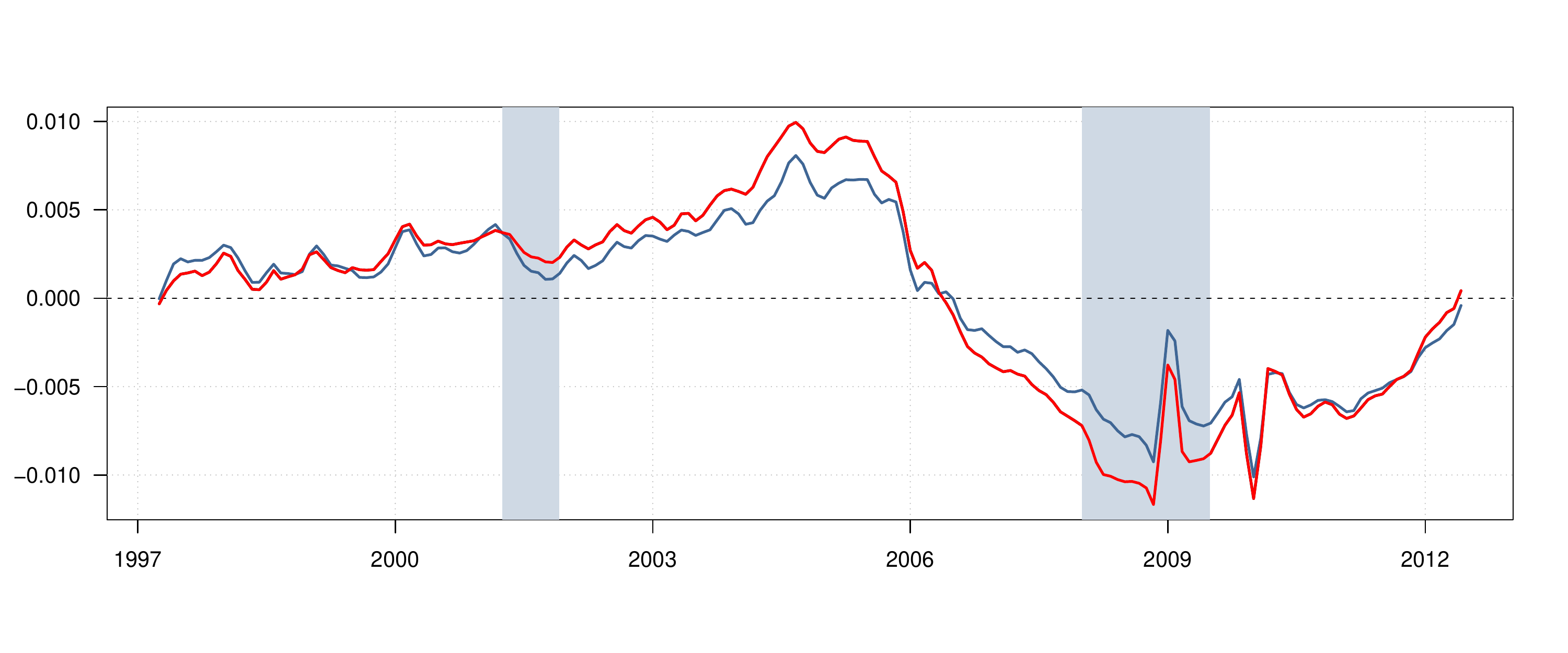}
\caption{Co-movement of the negative latent dynamic factor and national house prices over time.}\label{fig:dynamic-factor}
\caption*{\footnotesize{\noindent\textit{Notes}: The solid red line denotes the negative latent factor, i.e. $-F_t$, the solid blue line the national housing prices, calculated as mean of the individual regions. The dashed black line refers to the zero line, while the grey shaded areas represent the recessions by the Business Cycle Dating Committee of the National Bureau of Economic Research (www.nber.org). Sample period: 1997:04 -- 2012:06. Vertical axis: growth rates. Front axis: months.}}
\end{figure}

\cref{fig:dynamic-factor} shows the evolution of the negative latent factor (in solid red) and provides evidence that the common factor co-moves with the average growth rate of housing prices (in solid blue, calculated using the arithmetic mean of the individual regional housing prices) nearly perfectly. The figure illustrates that during the 2001 recession, housing price declines have been mild, while being substantial during the Great Recession, with large variations across space. It is worth noting that home prices fell the most during the late 2000s in regions with the largest declines in economic activity \citep{beraja2017}.

\begin{figure}[!t]
\centering
\includegraphics[width=\textwidth,trim={2.8cm 0 2.8cm 0},clip]{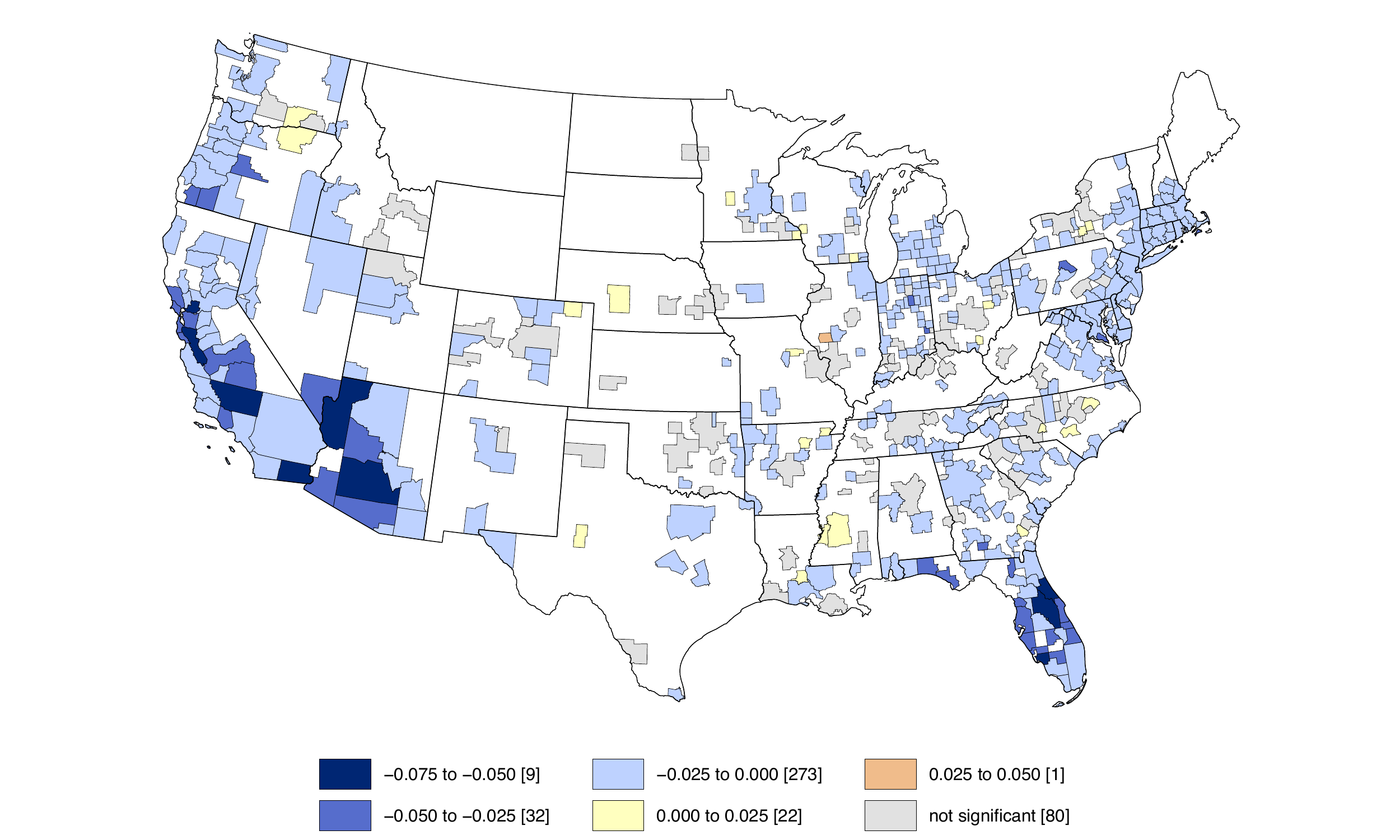}
\caption{Region-specific factor loadings.}\label{fig:regional-loadings}
\caption*{\footnotesize{\noindent\textit{Notes}: Visualization is based on a classification scheme with equal-interval breaks. Number of regions allocated to the classes in squared brackets. Thinner lines denote the boundaries of the 417 regions, while thicker lines represent US state boundaries. Results are based on 10,000 posterior draws. Sample period: 1997:04 -- 2012:06. For the list of regions see \cref{app:regions}.}}
\end{figure}

While \autoref{fig:dynamic-factor} provides intuition on the shape of the latent housing factor, the question on how individual regions are linked to it still needs to be addressed. For this purpose, \autoref{fig:regional-loadings} reports the posterior mean of the region-specific factor loadings in form of a geographic map in which thinner lines denote the boundaries of the regions, while thicker lines signify US state boundaries. Visualization is based on a classification scheme with equal-interval breaks. We see that the great majority of regions exhibit negative loadings, and only 22 regions show positive values. Eighty regions have zero loadings or loadings where the 16th and 84th credible sets (68 percent posterior coverage) of the respective posterior distributions include zero. The pattern of factor loadings, evidenced by the map, indicates that the latent factor is largely driven by regions located in California, Arizona and Florida. Regions in the rest of the country, with loadings being either small in absolute terms or not significantly different from zero, tend to play only a minor role in shaping national housing prices.

\subsection{Impulse responses of housing prices}
\begin{figure}[!htbp]
\centering
\includegraphics[width=.329\textwidth]{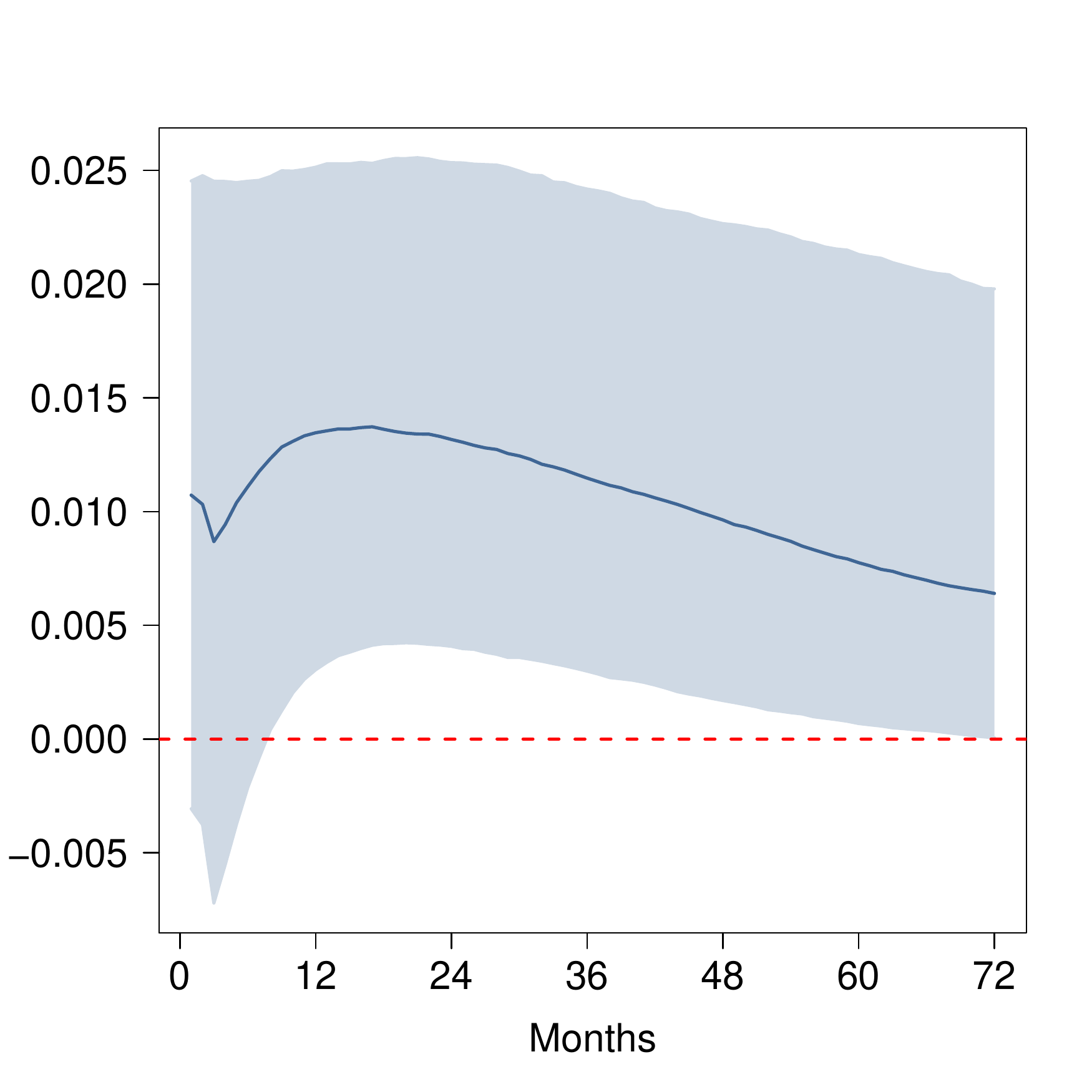}
\caption{The impulse response function of the latent factor, following a monetary policy shock.}\label{fig:irf-factor}
\caption*{\footnotesize{\textit{Notes}: The solid blue line denotes the median response, the dashed red line the zero line, and the shaded bands (in light blue) the 68 percent posterior coverage interval. Results are based on 10,000 posterior draws. Vertical axis: percentage points. Front axis: months after impact.}}
\end{figure}

\cref{fig:irf-factor} displays the impulse response function of the latent factor over 72 months after impact to an expansionary monetary policy shock. The latent factor reacts positively after the shock; however, the posterior coverage interval includes zero for the first few months. This is consistent with economic theory which suggests that as the costs of financing a home purchase decrease, the demand for housing increases and as a result, real housing prices increase.

\begin{figure}[!t]
\includegraphics[width=\textwidth,trim={2.8cm 0 2.8cm 0},clip]{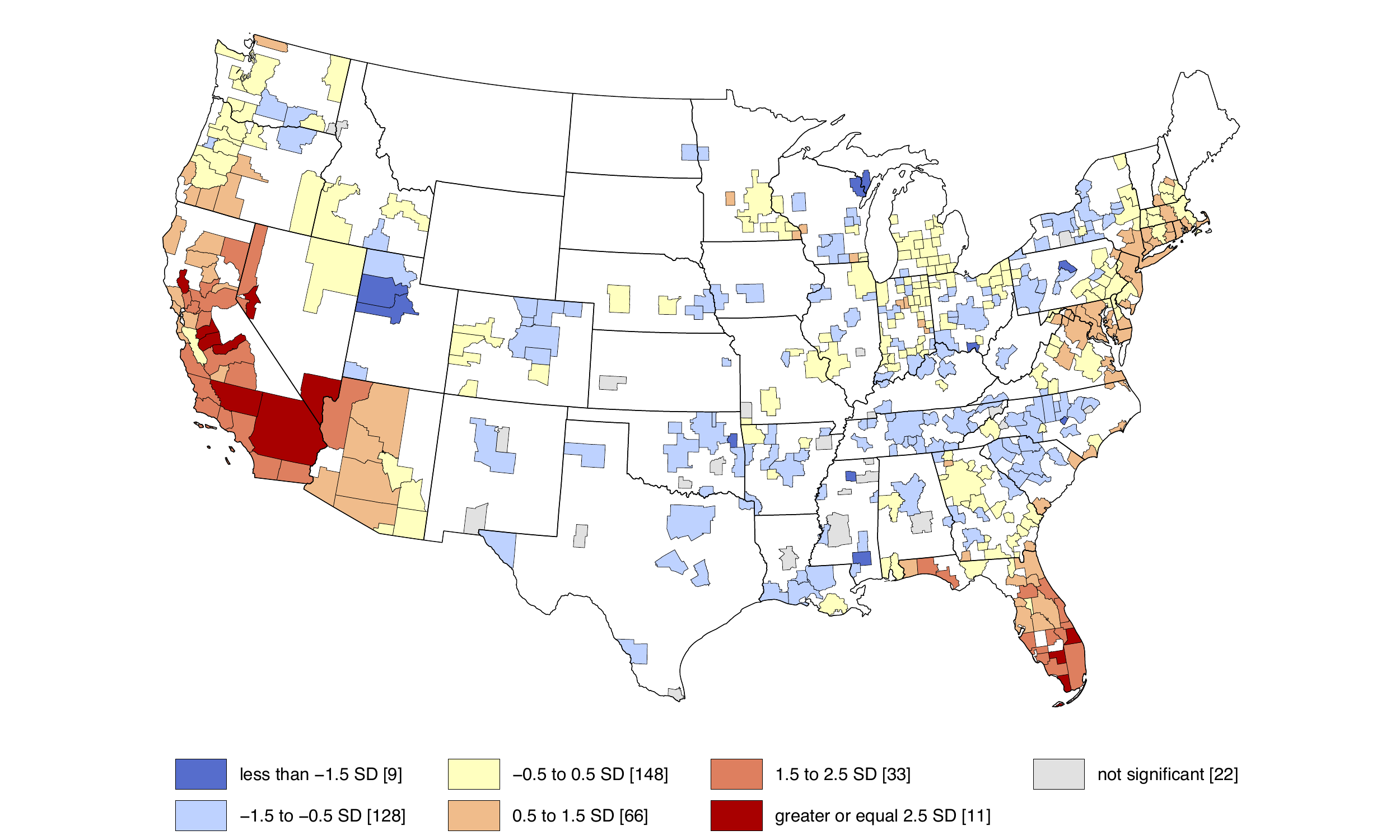}
\caption{Cumulative responses of regional housing prices to a monetary policy shock.}
\label{fig:regional-responses}
\caption*{\footnotesize{\textit{Notes}: Visualization is based on a classification scheme that generates breaks in standard deviation measures (SD $= 2.98$) above or below the mean of $3.43$. Number of regions allocated to the classes in squared brackets. The responses based on 10,000 posterior draws have been accumulated. Thinner lines denote the boundaries of the 417 regions, while thicker lines represent US state boundaries. Sample period: 1997:04 -- 2012:06. For the list of regions see \cref{app:regions}.}}
\end{figure}

Housing price responses, cumulated over the time horizon of six years, are displayed in \autoref{fig:regional-responses}.\footnote{The quantitative and qualitative nature of the results is robust to an alternative identification scheme, in which sign restrictions have been employed (for the results see \cref{app:robustness}).} The results are presented in form of a geographic map with a classification scheme that generates class breaks in standard deviation measures (SD $= 2.98$) above and below the mean of $3.43$. Again thinner lines denote the boundaries of the metropolitan regions and thicker lines those of US states.

Five points are worth noting here. \textit{First}, cumulative regional housing price effects vary substantially over space, with size and modest sign differences among the regions. Some few regions in Utah, New Mexico, Kansas, Oklahoma, Mississippi and West Virginia, but also in Louisiana and North Carolina show no significant impact or even negative cumulative responses. In more than 97 percent of the regions, however, the cumulative response of housing prices is positive. \textit{Second}, this heterogeneity may be due to varying sensitivity of housing to interest rates across space, and regional differences in housing markets, such as supply and demand elasticities \citep{fratatoni2003monetary}. For example, supply elasticities are relatively low on the East and West Coasts, but higher in the South and Southwest parts of the US. \textit{Third}, the largest cumulative effects can be observed in states on both the East and West Coasts, notably Riverside, Madera, Merced and Bakersfield in California, Miami-Fort Lauderdale and Key West in Florida, but also Las Vegas and Fernley in Nevada. These regions, expecially those in California, seem to play an important role in shaping the movement of the US housing price following a monetary shock.

\textit{Fourth}, the regions in the East North Central states (as defined by the Census Bureau), but also in Georgia and Massachusetts have cumulative home price responses that resemble the mean response of the US regions within a 0.5 standard deviation band from the mean. Prominent examples include Atlanta ($2.97$), Boston ($3.76$) and Chicago ($3.88$). \textit{Fifth} and finally, cumulative responses tend to be similar within states and adjacent regions in neighboring states. Looking at the map, this spatial autocorrelation phenomenon becomes particularly evident in the case of the Californian regions. This is most likely due to the importance of new house construction industries in California, along with the spatial influence the Californian housing market has on regions in neighboring states, especially Nevada and Arizona.

\begin{figure}[!t]
\centering
\begin{subfigure}{.329\textwidth}
\subcaption{Riverside, CA}\label{fig:irf_select5}
\includegraphics[width=\textwidth]{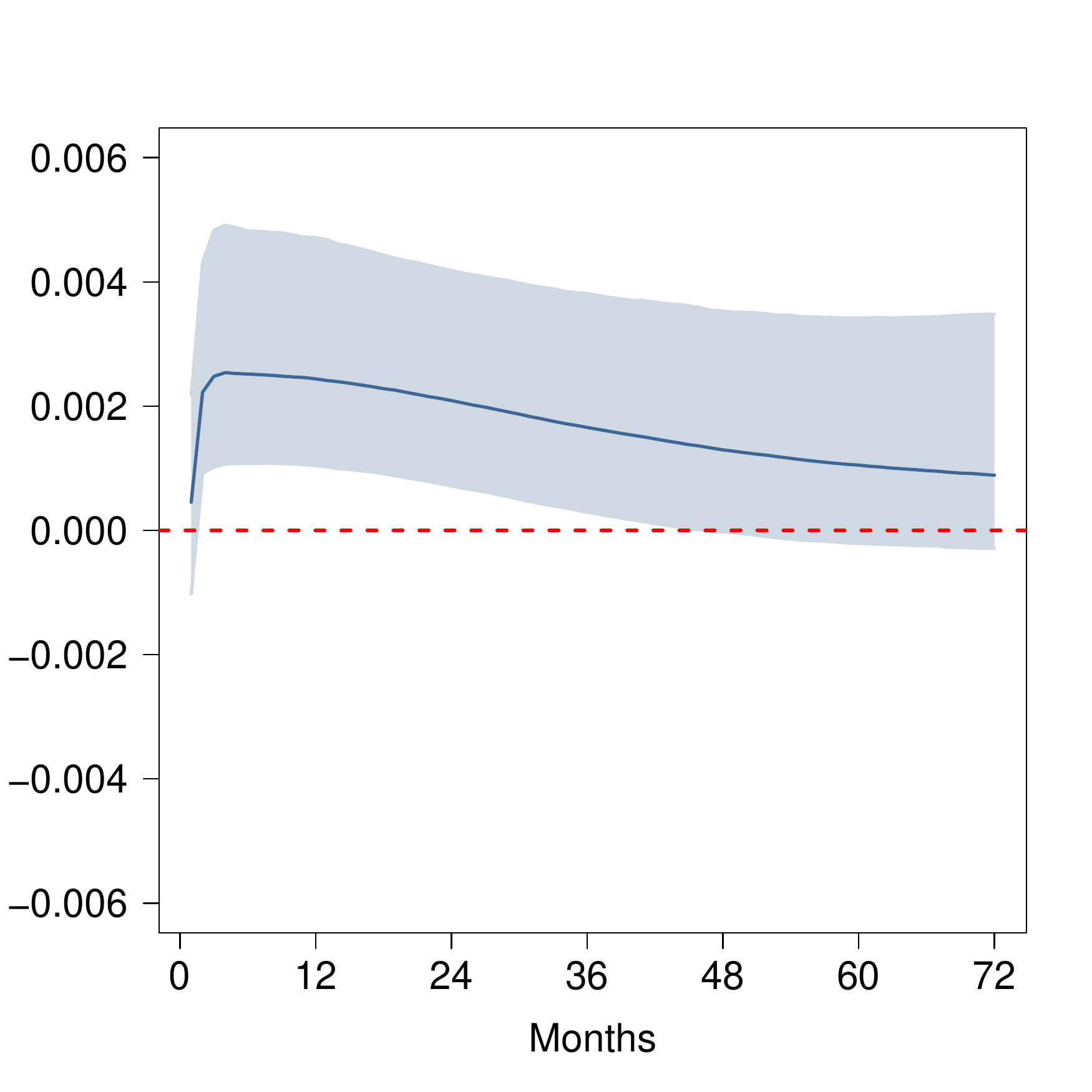}
\end{subfigure}
\begin{subfigure}{.329\textwidth}
\subcaption{Salt Lake City, UT}\label{fig:irf_select7}
\includegraphics[width=\textwidth]{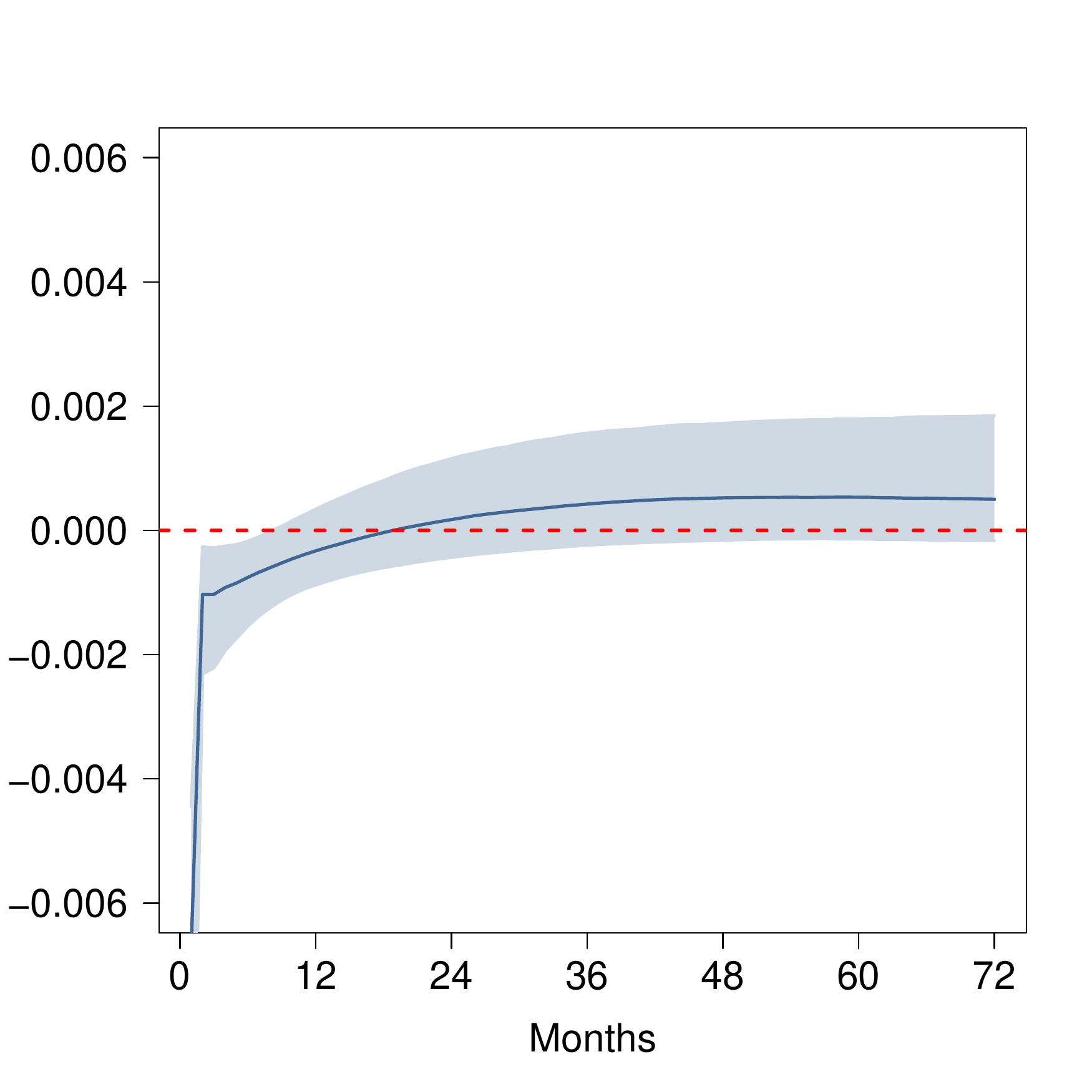}
\end{subfigure}
\begin{subfigure}{.329\textwidth}
\subcaption{Chicago, IL}\label{fig:irf_select2}
\includegraphics[width=\textwidth]{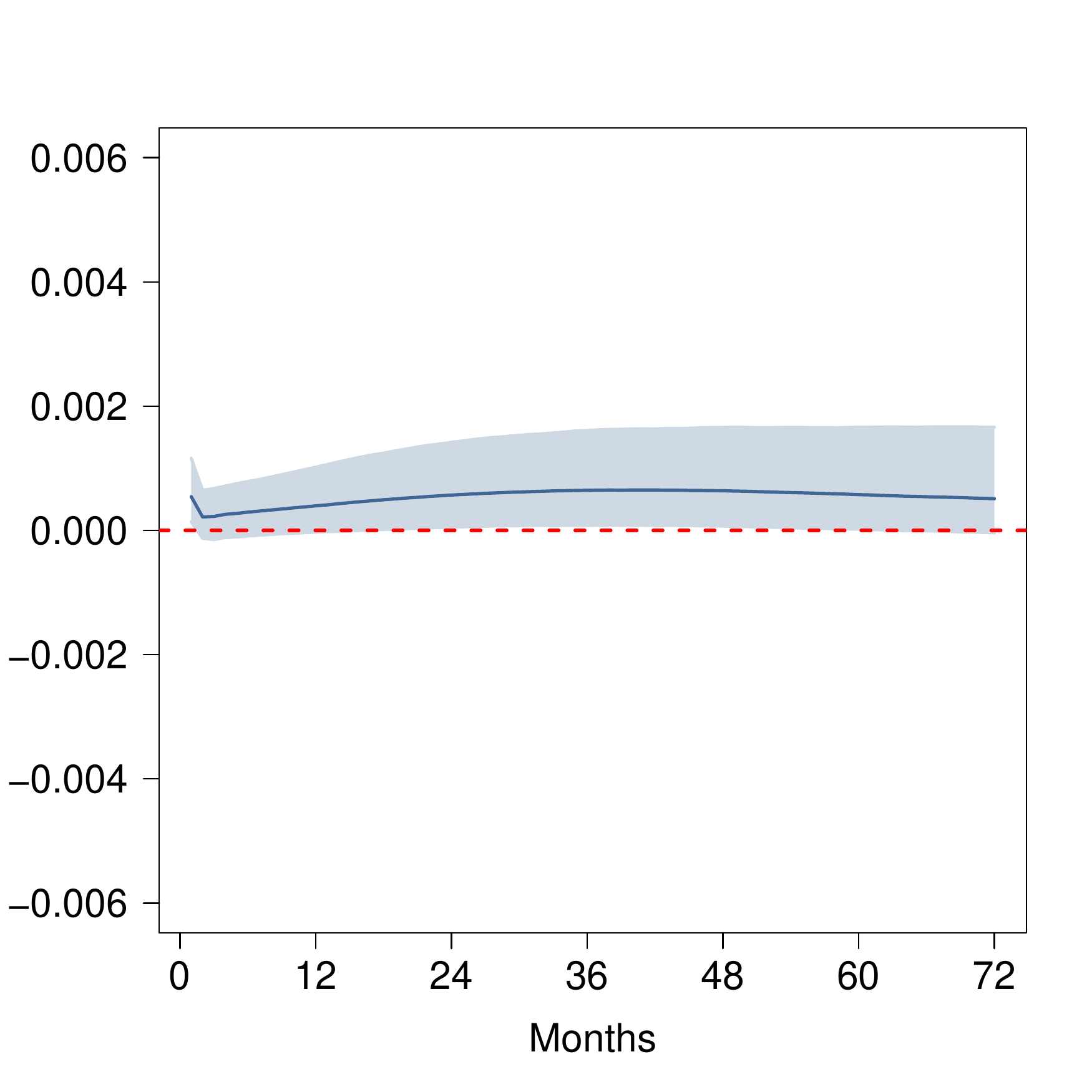}
\end{subfigure}\\
\hfill
\begin{subfigure}{.329\textwidth}
\subcaption{Miami-Fort Lauderdale, FL}\label{fig:irf_select6}
\includegraphics[width=\textwidth]{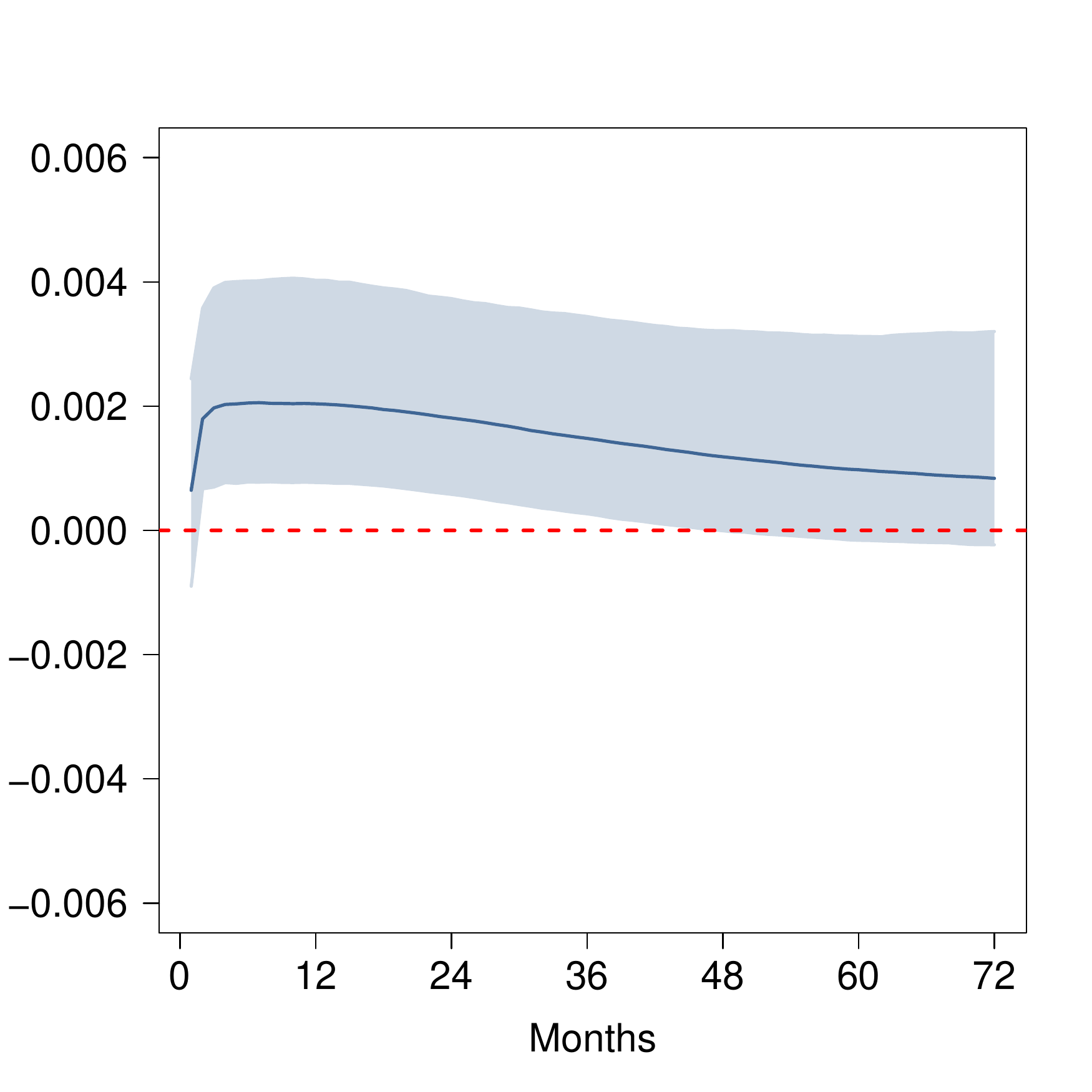}
\end{subfigure}
\begin{subfigure}{.329\textwidth}
\subcaption{Hickory, NC}\label{fig:irf_select8}
\includegraphics[width=\textwidth]{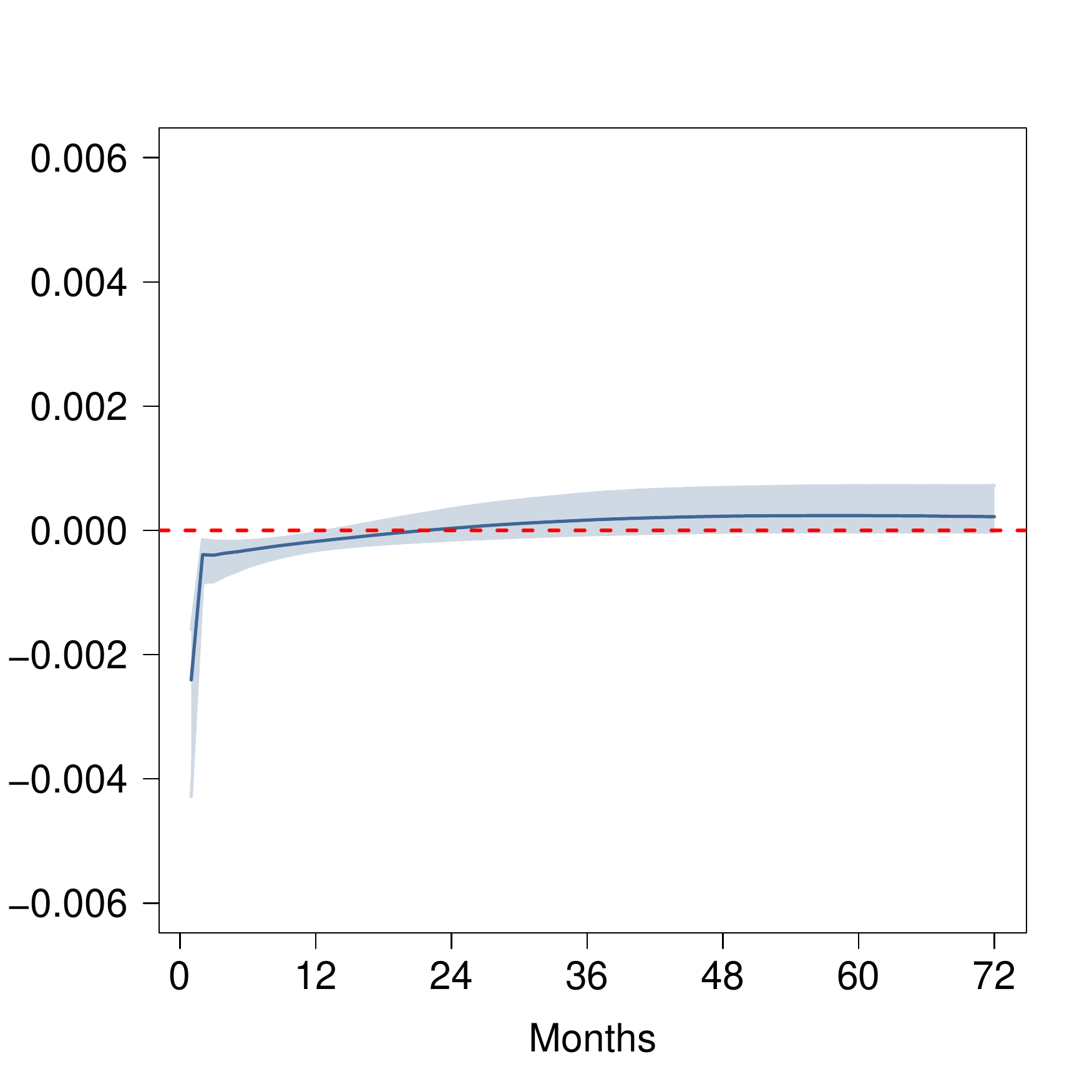}
\end{subfigure}
\begin{subfigure}{.329\textwidth}
\subcaption{Boston, MA}\label{fig:irf_select1}
\includegraphics[width=\textwidth]{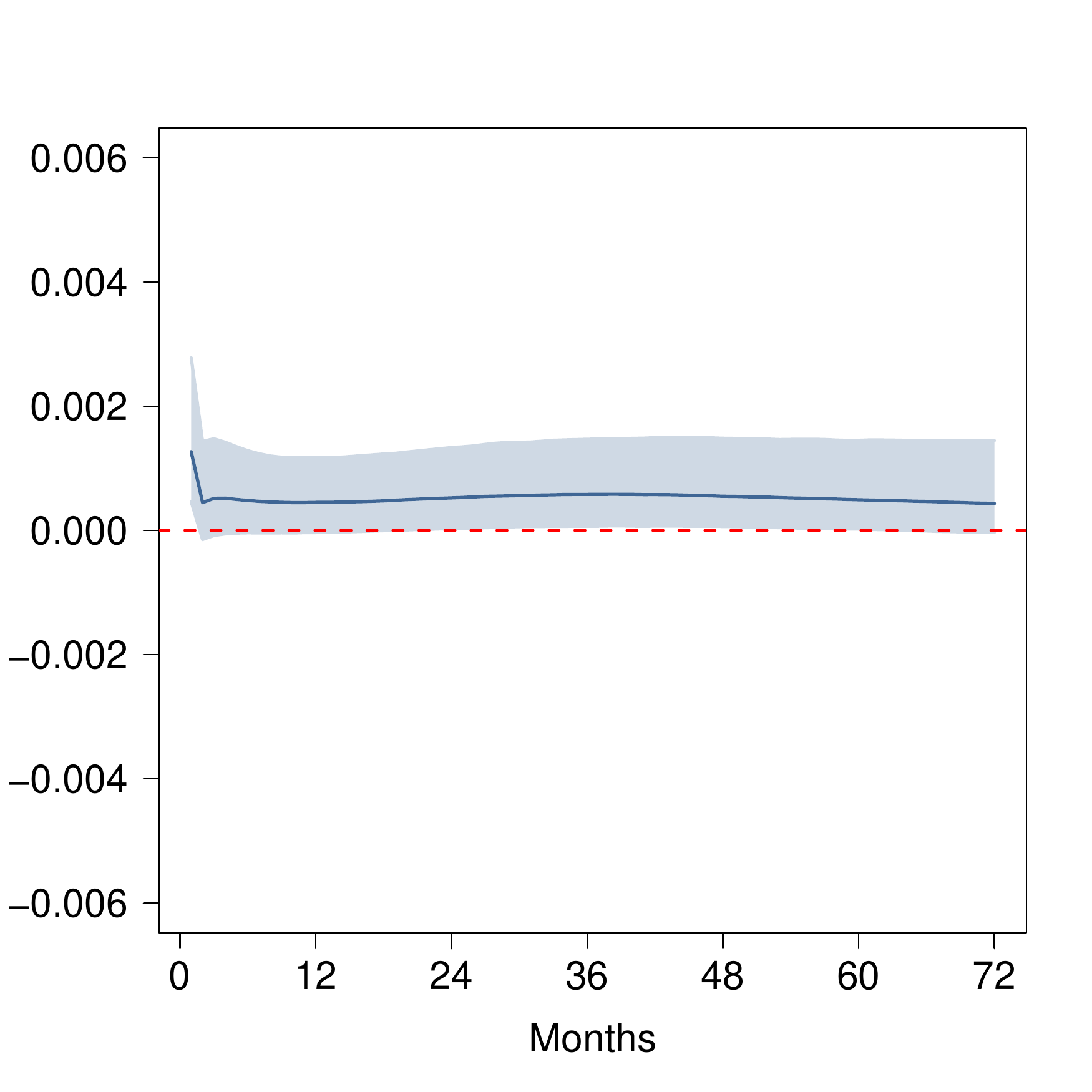}
\end{subfigure}

\caption{Impulse response functions of selected individual regions' housing prices.}\label{fig:IRF_selected}
\caption*{\footnotesize{\textit{Notes}: The solid blue line denotes the median response, the dashed red line the zero line, and the shaded bands (in light blue) the 68 percent posterior coverage interval. Riverside and Miami-Fort Lauderdale represent examples of metropolitan regions with large positive responses, Salt Lake City and Hickory those with negative responses, while Chicago and Boston closely resemble the mean response of the US regions. Results are based on 10,000 posterior draws. Sample period: 1997:04 -- 2012:06. Vertical axis: percentage points. Front axis: months after impact.}}
\end{figure}

For reasons of space limits we cannot present the 417 impulse response functions of the individual metropolitan regions, but report those of six regions in \autoref{fig:IRF_selected}. We pick these regions to show examples for metropolitan regions with larger positive cumulative responses (Riverside and Miami-Fort Lauderdale) and those with negative cumulative responses (Salt Lake City and Hickory). Recall that there are only eleven regions belonging to this latter category. For comparison, we also display the impulse response function of two regions that closely resemble the mean response of US regions (Chicago and Boston). Again, the solid blue line denotes the median response and the shaded areas (in light blue) the 68th posterior coverage intervals.

\cref{fig:IRF_selected} reveals profound differences in dynamic responses between the three regional categories, especially in shape and duration of effects. The differences within the categories tend to be rather small. In case of the first category, represented by the metropolitan regions Riverside and Miami-Fort Lauderdale, an expansionary monetary policy shock generates a significant increase of housing prices. This level remains stable and significant in the short-run, before fading away after approximately three years. The charts of Salt Lake City and Hickory, examples of the second regional category, show the housing price responses to fall strongly immediately, and these effects remain significantly negative for less than one year after impact. The response pattern of housing prices in Chicago and Boston is different. The effects are small in size, and hardly different from zero, with the exception of weakly significant effects between the third and fourth year after impact.

\section{Closing remarks}
This paper has examined the relationship between monetary policy and the US housing market, focusing on monetary policy shocks. The analysis is based on a Bayesian FAVAR model where monetary policy shocks are identified using high-frequency surprises around policy announcements as external instruments. Bayesian model estimation uses Gibbs sampling with Normal-Gamma shrinkage priors for both the autoregressive coefficients and factor loadings, relying on a panel of monthly time series for a set of 417 regions that range from 1997:04 to 2012:06.

The main findings of our analysis can be summarized as follows. The results provide empirical evidence that metropolitan regions react differently to an expansionary monetary policy shock, revealing magnitude and duration differences, and pointing to some modest sign differences, in additon. The extent and nature of regional heterogeneity are consistent with \citet{fratatoni2003monetary} who report impulse responses for house prices in 27 US metropolitan regions. Since our sample of regions covers the whole US (except Alaska, Maine, South Dakota and Wyoming) rather than only 16 US states, we find considerably greater regional heterogeneity in the results. In line with theory, the great majority of regions exhibit positive home price responses. The largest positive effects, cumulated over the time horizon of six years, can be observed in regions located in states on both the East and West Coasts, notably California, Arizona and Florida, and in Nevada. Impulse responses of regions tend to be similar within states and adjacent regions in neighboring states, evidenced by a high degree of spatial dependence among the impulse responses, as measured in terms of Moran's \textit{I} statistic.

Finally, it is worth noting that our analysis is confined to a linear setting, implying the underlying transmission mechanism to be constant over time. This assumption simplifies the analysis, but may be overly simplistic in turbulent economic times such as the collapse of the housing market around the Great Recesssion. Hence, an extension of the linear setting to allow for non-linearities -- in the spirit of \citet{huber2018} -- might be a promising avenue for future research.

\small
\newpage
\singlespacing
\bibliographystyle{fischer}
\bibliography{favar,mpShocks,additional}
\addcontentsline{toc}{section}{References}
\newpage
\onehalfspacing\normalsize

\begin{appendices}\crefalias{section}{appsec}
\setcounter{equation}{0}
\renewcommand\theequation{A.\arabic{equation}}
\section{The MCMC algorithm}
\label{app:mcmc}
We estimate the model by running an MCMC algorithm. The full conditional posterior distributions are available in closed form implying that we can apply Gibbs sampling to obtain draws from the joint posterior distribution. More specifically, our MCMC algorithm involves the following steps:

\singlespacing\begin{enumerate}[label=(\roman*)]
\setlength\itemsep{0em}
\item Simulate the VAR coefficients $a_j~(j = 1,\hdots,J)$ conditional on the factors and remaining model parameters from a multivariate Gaussian distribution that takes a standard form \citep[see, for instance,][for further information]{george2008bayesian}.
\item Simulate the latent factors $\bm{F}_t~(t = 1, \hdots, T)$ by using forward filtering backward sampling \citep{carter1994gibbs, fruhwirth1994data}.
\item The error variance-covariance matrix $\bm{\Sigma}_u$ is simulated from an inverted Wishart posterior distribution with degrees of freedom equal to $\nu = v + T$ and scaling matrix equal to $\bm{P}=\sum_{t=1}^T (\bm{y}_t-\bm{A} \bm{x}_t)'(\bm{y}_t-\bm{A} \bm{x}_t) + \bm{\ubar{\Sigma}}$.
\item Simulate the factor loadings $\lambda_\ell~(\ell = 1, \hdots, L)$ from Gaussian posteriors (conditioned on the remaining parameters and the latent factors) by running a sequence of $(R-S)$ unrelated regression models.
\item The measurement error variances $\sigma_r^2$ for $r = S+1, \dots, R$ are simulated independently from an inverse Gamma distribution  $\sigma_r^2| \Xi \sim \mathcal{G}^{-1}(\alpha_r, \beta_r)$ with $\alpha_r = \frac{1}{2} T + e_0$ and  $\beta_r = \frac{1}{2} \sum_{t=1}^T (H_{rt} - \bm{\Lambda}^F_{r \bullet} \bm{F}_t - \bm{\Lambda}^M_{r \bullet} \bm{M}_t)^2+e_1$. The notation $\bm{\Lambda}^F_{r \bullet}$ indicates that the $r$th row of the matrix concerned is selected and $\Xi$ stands for conditioning on the remaining parameters and the data.
\item Simulate $\tau^2_{a j}~(j=1,\dots,J)$ from a generalized inverted Gaussian distributed posterior distribution with
\begin{equation}
\tau^2_{a j}|\Xi \sim \mathcal{GIG}\left(\vartheta_a - \frac{1}{2}, a_j^2, \vartheta_a \xi_a\right).
\end{equation}
\item Draw $\xi_a$ from a Gamma distributed posterior given by
\begin{equation}
\xi_a|\Xi \sim \mathcal{G}\left(c_0 + \vartheta_a J, c_1 + \frac{1}{2} \vartheta_a \sum_{\ell=1}^L \tau^2_{a \ell}\right).
\end{equation}
\item Simulate the posterior of $\tau^2_{\lambda \ell}~(\ell=1,\dots,L)$ from a generalized inverted Gaussian distribution,
\begin{equation}
\tau^2_{\lambda \ell}|\Xi \sim \mathcal{GIG}\left(\vartheta_\lambda-\frac{1}{2}, \lambda_\ell^2, \vartheta_\lambda \xi_\lambda\right).
\end{equation}
\item Finally, the global shrinkage parameter $\xi_\lambda$ associated with the prior on the factor loadings is simulated from a Gamma distribution,
\begin{equation}
\xi_\lambda|\Xi \sim \mathcal{G}\left(d_0 +\vartheta_\lambda L, d_1 + \frac{1}{2}\vartheta_\lambda \sum_{\ell=1}^L \tau^2_{\lambda \ell}\right).
\end{equation}
\end{enumerate}
\onehalfspacing
Steps described above are iterated for 20,000 cycles, where we discard the first 10,000 draws as burn-in.
\newpage

\section{Regions used in the study}
\label{app:regions}
Regions in this study are defined as core-based statistical areas (CBSA) that -- by definition of the United States Office of Management and Budget -- are based on the concept of a core area of at least 10,000 population, plus adjacent counties having at least 25 percent of employed residents of the county who work in the core area. Core-based statistical areas may be categorized as being either metropolitan or micropolitan. The 917 core-based statistical areas include 381 metropolitan statistical areas which have an urban core population of at least 50,000, and 536 micropolitan statistical areas which have an urban core population of at least 10,000 but less than 50,000. In this study we use 264 metropolitan and 153 micropolitan statistical areas, due to limited availability of data. These 417 regions, briefly termed metropolitan regions in this paper, represent all US states except Alaska, Maine, South Dakota and Wyoming.

\singlespacing
\footnotesize
\begin{longtable}{p{.2\textwidth} p{.75\textwidth}}
\caption{The list of metropolitan statistical areas used.}\\
\toprule
\textbf{State} & \textbf{Region} \\ 
  \midrule\endfirsthead

 & \multicolumn{1}{r}{\textit{ctd.}}\\
 \midrule
\endhead

\midrule
\endfoot

\endlastfoot
  Alabama & Birmingham, Daphne, Mobile, Montgomery, Tuscaloosa \\ 
  Arizona & Flagstaff, Lake Havasu City, Phoenix, Prescott, Sierra Vista, Tucson, Yuma \\ 
  Arkansas & Fayetteville, Fort Smith, Hot Springs, Jonesboro, Little Rock \\ 
  California & Bakersfield, Chico, El Centro, Fresno, Hanford, Los Angeles-Long Beach-Anaheim, Madera, Merced, Modesto, Napa, Redding, Riverside, Sacramento, Salinas, San Diego, San Francisco, San Jose, San Luis Obispo, Santa Cruz, Santa Maria-Santa Barbara, Santa Rosa, Stockton, Vallejo, Ventura, Visalia, Yuba City \\ 
  Colorado & Boulder, Colorado Springs, Denver, Fort Collins, Grand Junction, Greeley, Pueblo \\ 
  Connecticut & Hartford, New Haven, New London, Stamford \\ 
  Delaware & Dover \\ 
  District of Columbia & Washington \\ 
  Florida & Crestview-Fort Walton Beach-Destin, Daytona Beach, Fort Myers, Gainesville, Homosassa Springs, Jacksonville, Lakeland, Melbourne, Miami-Fort Lauderdale, Naples, North Port-Sarasota-Bradenton, Ocala, Orlando, Panama City, Pensacola, Port St. Lucie, Punta Gorda, Sebring, Tallahassee, Tampa, The Villages, Vero Beach \\ 
  Georgia & Albany, Athens, Atlanta, Augusta, Columbus, Dalton, Gainesville, Hinesville, Macon, Savannah, Valdosta, Warner Robins \\ 
  Hawaii & Kahului, Urban Honolulu \\ 
  Idaho & Boise City, Idaho Falls, Lewiston \\ 
  Illinois & Bloomington, Chicago, Davenport, Kankakee, Springfield \\ 
  Indiana & Bloomington, Elkhart, Evansville, Fort Wayne, Lafayette-West Lafayette, Muncie, South Bend, Terre Haute \\ 
  Iowa & Des Moines \\ 
  Kansas & Lawrence \\ 
  Kentucky & Lexington, Louisville-Jefferson County \\ 
  Louisiana & Alexandria, Baton Rouge, Houma, Lafayette, Lake Charles \\ 
  Nebraska & Grand Island, Lincoln, Omaha \\ 
  Nevada & Las Vegas, Reno \\ 
  New Hampshire & Manchester \\ 
  New Jersey & Ocean City, Trenton, Vineland \\ 
  New Mexico & Albuquerque, Las Cruces, Santa Fe \\ 
  New York & Albany, Binghamton, Elmira, Glens Falls, Ithaca, Kingston, New York, Rochester, Syracuse, Watertown \\ 
  North Carolina & Asheville, Burlington, Charlotte, Durham, Fayetteville, Greensboro, Hickory, Raleigh, Rocky Mount, Wilmington, Winston-Salem \\ 
  North Dakota & Fargo \\ 
  Ohio & Akron, Canton, Cincinnati, Cleveland, Columbus, Dayton, Lima, Springfield, Toledo, Youngstown \\ 
  Oklahoma & Oklahoma City, Tulsa \\ 
  Oregon & Albany, Bend, Corvallis, Eugene, Grants Pass, Medford, Portland, Salem \\ 
  Maryland & Baltimore, California-Lexington Park, Cumberland, Hagerstown, Salisbury \\ 
  Massachusetts & Boston, Cape Cod, Pittsfield, Springfield, Worcester \\ 
  Michigan & Ann Arbor, Battle Creek, Bay City, Grand Rapids, Jackson, Lansing, Midland, Monroe, Muskegon, Saginaw \\ 
  Minnesota & Mankato, Minneapolis-St Paul, Rochester \\ 
  Mississippi & Hattiesburg, Jackson \\ 
  Missouri & Columbia, Joplin, Springfield, St. Louis \\ 
  Pennsylvania & Allentown, Altoona, Erie, Harrisburg, Lancaster, Philadelphia, Pittsburgh, Reading, Scranton, State College, York \\ 
  Rhode Island & Providence \\ 
  South Carolina & Columbia, Florence, Greenville, Hilton Head Island, Myrtle Beach, Spartanburg \\ 
  Tennessee & Chattanooga, Clarksville, Cleveland, Jackson, Johnson City, Kingsport, Knoxville, Nashville \\ 
  Texas & Amarillo, Brownsville, College Station, Dallas-Fort Worth, El Paso, Killeen, Laredo, Midland, Texarkana \\ 
  Utah & Ogden, Provo, Salt Lake City, St. George \\ 
  Virginia & Charlottesville, Harrisonburg, Richmond, Roanoke, Staunton, Virginia Beach, Winchester \\ 
  Washington & Bellingham, Kennewick, Longview, Olympia, Seattle, Spokane, Walla Walla, Yakima \\ 
  West Virginia & Charleston \\ 
  Wisconsin & Appleton, Eau Claire, Fond du Lac, Janesville, La Crosse, Madison, Oshkosh, Racine \\ 
\bottomrule
\end{longtable}

\footnotesize
\begin{longtable}{p{.2\textwidth} p{.75\textwidth}}
\caption{The list of micropolitan statistical areas used.}\\
\toprule
\textbf{State} & \textbf{Region} \\ 
  \midrule\endfirsthead

 & \multicolumn{1}{r}{\textit{ctd.}}\\
 \midrule
\endhead

\midrule
\endfoot

\endlastfoot
  Arizona & Nogales, Payson, Safford \\ 
  Arkansas & Batesville, Harrison, Paragould, Russellville, Searcy \\ 
  California & Clearlake, Eureka, Red Bluff, Susanville, Truckee \\ 
  Colorado & Durango, Glenwood Springs, Montrose, Sterling \\ 
  Connecticut & Torrington \\ 
  Florida & Clewiston, Key West, Lake City, Okeechobee, Palatka \\ 
  Georgia & Bainbridge, Calhoun, Cedartown, Dublin, Jesup, Moultrie, St. Marys, Thomaston, Tifton, Vidalia, Waycross \\ 
  Hawaii & Hilo \\ 
  Idaho & Burley \\ 
  Illinois & Effingham, Jacksonville \\ 
  Indiana & Angola, Auburn, Bedford, Connersville, Crawfordsville, Decatur, Frankfort, Greensburg, Huntington, Jasper, Kendallville, Logansport, Madison, Marion, New Castle, North Vernon, Peru, Plymouth, Richmond, Seymour, Vincennes, Wabash, Warsaw, Washington \\ 
  Kansas & Garden City \\ 
  Kentucky & Danville, Murray \\ 
  Louisiana & Opelousas \\ 
  Nebraska & North Platte \\ 
  Nevada & Elko, Fernley, Gardnerville Ranchos \\ 
  New Hampshire & Concord, Keene, Laconia \\ 
  New York & Amsterdam, Batavia, Corning, Cortland, Gloversville, Hudson, Olean, Oneonta, Plattsburgh, Seneca Falls \\ 
  North Carolina & Albemarle, Morehead City, Sanford, Wilson \\ 
  Ohio & Ashtabula, Coshocton, Defiance, Findlay, Jackson, New Philadelphia, Portsmouth, Sandusky, Urbana, Wooster \\ 
  Oklahoma & Ardmore, Bartlesville, Duncan, Durant, Enid, McAlester, Tahlequah \\ 
  Oregon & Coos Bay, Hermiston-Pendleton, Klamath Falls, Ontario, Roseburg, The Dalles \\ 
  Maryland & Cambridge, Easton \\ 
  Massachusetts & Greenfield Town, Vineyard Haven \\ 
  Michigan & Adrian, Hillsdale, Holland, Ionia, Ludington, Owosso \\ 
  Minnesota & Owatonna, Willmar, Winona \\ 
  Mississippi & Cleveland, Columbus, Corinth, Grenada, Laurel, Oxford, Picayune, Tupelo, Vicksburg \\ 
  Missouri & Mexico \\ 
  Pennsylvania & Indiana, Lock Haven, Oil City, Pottsville \\ 
  South Carolina & Orangeburg \\ 
  Tennessee & Cookeville, Lawrenceburg, Lewisburg, Martin, Paris, Sevierville, Shelbyville, Tullahoma \\ 
  Virginia & Danville, Martinsville \\ 
  Washington & Oak Harbor, Port Angeles, Shelton \\ 
  Wisconsin & Baraboo, Marinette, Whitewater \\ 
\bottomrule
\end{longtable}
\onehalfspacing\normalsize

\newpage
\section{Robustness check}
\label{app:robustness}
To assess the sensitivity of our results with respect to identification of the monetary policy shock, we use an alternative strategy based on contemporaneous sign restrictions \citep[see][]{uhlig2005effects,DEDOLA2007512}. Technical implementation is achieved by using the algorithm proposed in \citet{arias2014inference} that collapses to the procedure outlined in \citet{rubio2010structural} in the absence of zero restrictions. For each iteration of the MCMC algorithm we draw a rotation matrix and assess whether the following set of sign restrictions is satisfied. Consistent with economic common sense, output (measured in terms of the industrial production index), housing investment (measured in terms of housing starts) and consumer prices (measured in terms of the consumer price index) are bound to increase on impact. Moreover, we assume that the term-spread also widens on impact. Finally, consistent with the normalization adopted when using external instruments, we assume that the one-year yield declines. If this is the case, we keep the rotation matrix and store the associated structural coefficients, while if the sign restrictions are not met, we reject the draw and repeat the procedure.

\begin{figure}[!htbp]
\centering
\includegraphics[width=0.65\textwidth,trim={2.8cm 0 2.8cm 0},clip]{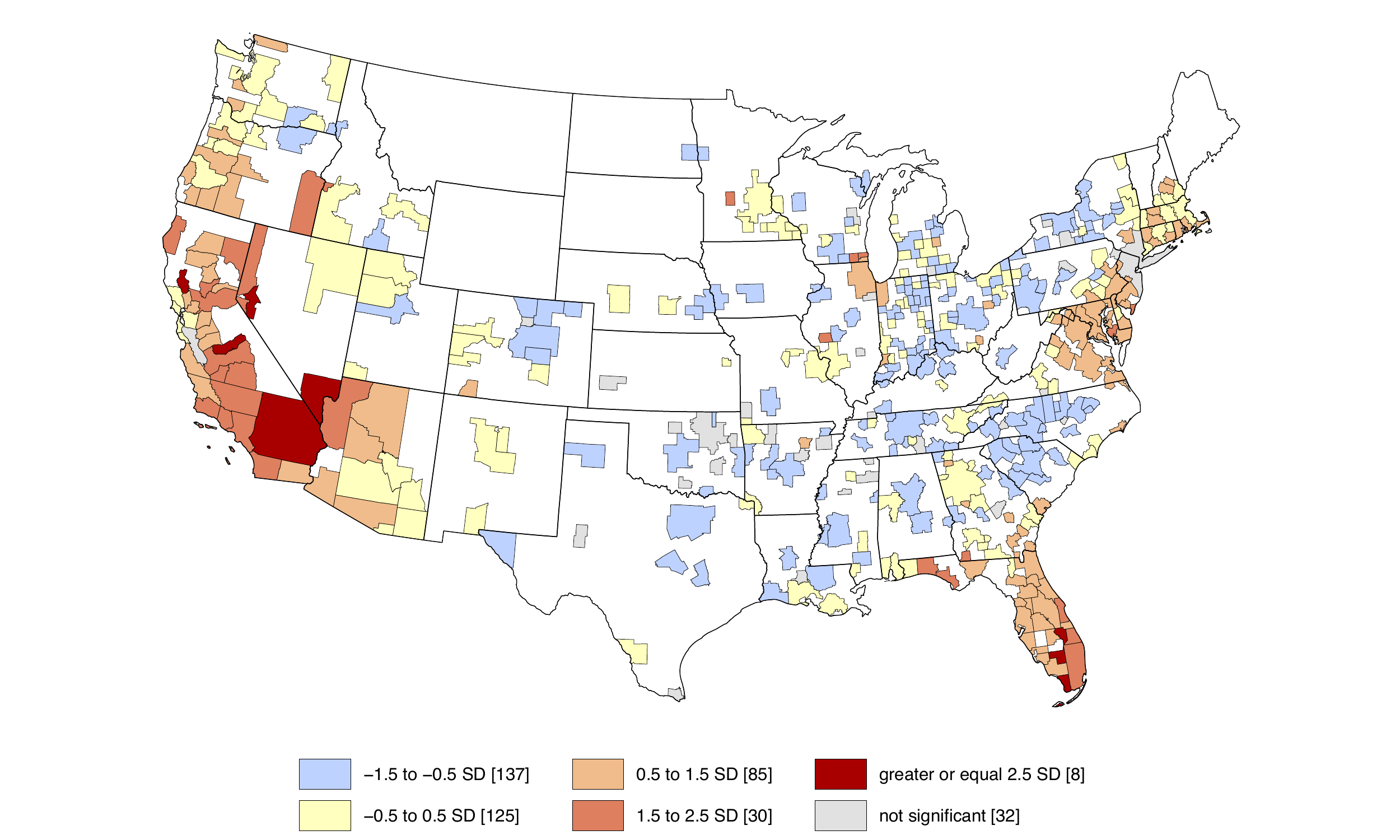}
\caption{Robustness check: Cumulative responses of regional housing prices to a monetary policy shock identified by means of sign restrictions.}\label{fig:regional-responses_rob}
\caption*{\footnotesize{\textit{Notes}: Visualization is based on a classification scheme that generates breaks in standard deviation measures. Number of regions allocated to the classes in squared brackets. The responses based on 10,000 posterior draws have been accumulated. Thinner lines denote the boundaries of the 417 regions, while thicker lines represent US state boundaries. Sample period: 1997:04--2012:06. For the list of regions see \cref{app:regions}.}}
\end{figure}

The results are displayed in form of a geographic map with a classification scheme that generates class breaks in standard deviation measures above and below the mean, see \autoref{fig:regional-responses_rob}. A comparison with \autoref{fig:regional-responses} provides evidence of the robustness of our results.

%

\end{appendices}
\end{document}